\documentclass[final, journal, twoside]{IEEEtran}

\ifCLASSINFOpdf
	\usepackage[pdftex]{graphicx}
\else
	\usepackage[dvips]{graphicx}
\fi
\graphicspath{{figure/}{src_figure/}}
\ifCLASSOPTIONcompsoc
  \usepackage[caption=false,font=footnotesize,labelfont=sf,textfont=sf]{subfig}
\else
  \usepackage[caption=false,font=footnotesize]{subfig}
\fi
\usepackage{epstopdf}
\usepackage{epsfig}
\DeclareGraphicsExtensions{.eps, .pdf}
\usepackage{color}		% color content
\usepackage{amsmath,amssymb,amsfonts}
\usepackage{cite}
\usepackage{url}		% hyperlinks
\usepackage{bm}			% bold type for equations
\usepackage{booktabs}
\usepackage{multirow}
\usepackage{balance}
\usepackage{algorithm}
\usepackage{algorithmic}
\usepackage[colorlinks, bookmarks, unicode]{hyperref}

\newcommand{\fro}{\mathrm{F}}
\newcommand{\fmat}[1]{\bm{\hat{#1}}}
\newcommand{\Rdim}[1]{\in\mathbb{R}^{#1}}

\begin{document}
\title{1D Probabilistic Undersampling Pattern Optimization for MR Image Reconstruction}
\author{Shengke~Xue,  Xinyu Jin, and Ruiliang~Bai
		\thanks{\emph{Corresponding author}: \emph{Ruiliang~Bai} (ruiliangbai@zju.edu.cn)}% <-this % stops a space
\thanks{S.~Xue and X. Jin are with the College of Information Science and Electronic Engineering, Zhejiang University, Hangzhou, P.~R.~China (e-mail: xueshengke@zju.edu.cn, jinxy@zju.edu.cn).}

\thanks{R.~Bai is with the Department of Physical Medicine and Rehabilitation of the Affiliated Sir Run Run Shaw Hospital AND Interdisciplinary Institute of Neuroscience and Technology, School of Medicine, Zhejiang University, Hangzhou, 310029, China. R.~Bai is with the Key Laboratory of Biomedical Engineering of Ministry of Education, College of Biomedical Engineering and Instrument Science, Zhejiang University, Hangzhou, 310029, China (e-mail: ruiliangbai@zju.edu.cn)}% <-this % stops a space

%\thanks{This work was supported by the National Natural Science Foundation of China (Grant No.~81873894) and the Natural Science Foundation of Zhejiang Province, China (Grant No.~LR20H180001).
%%the National Natural Science Foundation of China (No.~81873894).
%}

\thanks{Manuscript received *** **, 2020; revised *** **, 2020.}}

\markboth{IEEE Transactions on ******,~Vol.~**, No.~*, ****~2020}%
{Xue \MakeLowercase{\textit{et~al.}}: ***}

\maketitle

\begin{abstract}
	3D magnetic resonance imaging (MRI) provides higher image quality but is mainly limited by its long scanning time. \textit{k}-space undersampling was used to  accelerate the acquisition of MRI but always suffer from poorly reconstructed MR images. Recently, some studies 1) used effective undersampling patterns, or  2) designed deep neural networks to improve the image reconstruction from \textit{k}-space. However,  these studies considered  undersampling and reconstruction as two separate optimization strategies though they are entangled in theory. 
	In this study, we propose a cross-domain network for MR image reconstruction to  simultaneously obtain the optimal undersampling pattern (in \textit{k}-space under the Cartesian undersampling) and the reconstruction model. The reconstruction model is customized to the type of training data, by using an end-to-end learning strategy. 
	In this model, a 1D probabilistic undersampling layer was designed  to obtain the optimal undersampling pattern and its probability distribution in a differentiable way and a 1D inverse Fourier transform layer was implemented to connect the Fourier domain and the image domain during the forward pass and the backpropagation. 
	Finally, by training the network on 3D fully-sampled \textit{k}-space data and MR images of experimental data with the conventional Euclidean loss, we discover a universal relationship between the probability distribution of the optimal undersampling pattern and its corresponding sampling rate.
	Retrospective studies with experimental data show that the recovered MRI images using our 1D probabilistic undersampling pattern significantly outperform than those using the existing and start-of-the-art undersampling strategies in both qualitative and quantitative comparison. 
\end{abstract}


\IEEEpeerreviewmaketitle

\section{Introduction} \label{sec:introduction}

\IEEEPARstart{M}{agnetic} resonance imaging (MRI) is one of the crucial imaging tools for  clinical diagnosis, and fundamental research, due to its non-invasiveness, non-ionizing radiation, and outstanding contrast. Recently, 3D MRI acquisition is becoming popular due to its better performance in image quality and signal-to-noise ratio (SNR) than conventional 1D MRI acquisition. However, the long acquisition time is a primary drawback that limits the applications of 3D MRI. 
Parallel imaging relying on the multi-coil technology is useful to accelerate the MRI, but at the cost of hardware requirements. An alternative technique is compressed sensing (CS), because CS overcomes the limitation of the Nyquist-Shannon sampling law with the nature of redundancy in $ k $-space. To this end, recovering a high-quality MR image from its partial undersampled $ k $-space data by using CS becomes feasible without hardware modification. So far, a series of CS-MRI algorithms have been developed, including total variation (TV) \cite{osher2005an}, $k$-$t$ FOCUSS \cite{Jung2007Generalized}, nuclear norm \cite{Ongie2016Low_rank, Sun2015nonlocal} and low-rank completion \cite{candes2011robust, wright2009robust, Xue2018Low, Xue2018Robust} methods.

Dictionary learning \cite{Aharon2006K_SVD} based methods have been used in CS-MRI. However, they have to use numerous iterations and a great deal of time to converge, which can be computationally expensive. This indicates that they are not applicable for large-scale datasets. To this end, deep learning has shown its advantages in CS-MRI reconstruction tasks \cite{Wang2016Accelerating, Yang2017ADMM,Sun2017Compressed, Yang2018DAGAN, Mardani2017Deep, Hyun2018Deep, Quan2018Compressed, Schlemper2017Cascade, Zhang2017ISTA, pereira2016brain, shin2016deep, bahrami2016reconstruction}. They obviously reduce the running time of $ k $-space undersampling and reconstruction, and significantly improve the quality of final MR images. These studies design deep neural networks in either the image domain or the Fourier domain. Fig.~1 of \cite{Han2018kspace} provides a thoughtful review of deep learning in CS-MRI.

However, these kinds of methods just take the undersampled partial $ k $-space data as input, and do not consider the undersampling process as one part of the optimization framework. In other words, these studies usually adopt existing undersampling strategies, e.g., Gaussian undersampling or Poisson undersampling. However, it is still unclear whether such undersampling stratergy is the best for MRI reconstruction.

Since the model's performance is heavily dependent on the type of undersampling strategy,  there are some studies but few focus on optimizing the undersampling patterns to recover high-quality MR images in a data-driven fashion \cite{Gozcu2018LearningMRI, Huijben2020probabilisic_subsampling, Aggarwal2019J_MoDL, Weiss2020Joint}.
%, which will be introduced in Section~\ref{sec:related_work_optimize_undersampling_pattern}. 
However, they have several common drawbacks: 
\begin{enumerate}
	\item In application environments, sampling points are binary values and non-differentiable, so that gradient backpropagation cannot be directly applied in the training;
	
	\item Integrating the underlying property (rule) of $ k $-space data with the feature extraction capabilities of deep neural networks during training is difficult;
	
	\item Optimal undersampling pattern and its corresponding theoretical expression cannot be simultaneously discovered from limited amount of $ k $-space data.
\end{enumerate}

In this study, we propose a cross-domain network for 1D undersampling pattern optimization and MR image reconstruction in a retrospective data-driven fashion  and  under limited sampling rates (accelerating ratios). The major contributions of this study include:
\begin{itemize}
	\item We propose a 1D probabilistic undersampling layer  including differentiable probability and sampling matrices, to obtain the optimal undersampling pattern and analyze its corresponding probability distribution, which are customized to specific $ k $-space data.

	\item We propose a 1D inverse Fourier transform layer  to achieve the cross-domain scheme. Specifically, the forward pass from the Fourier domain to the image domain and the backpropagation in the inverse direction can be established. To this end, our method simultaneously learns the optimal undersampling pattern (in the Fourier domain) and the corresponding reconstruction model (in the image domain), in an end-to-end training strategy. 
	
	\item We discover a universal relationship between the probability distribution of our 1D undersampling pattern and its sampling rate, by training limited amount of 3D fully-sampled $ k $-space data and MR images and constrained to the conventional Euclidean loss. Thus, our results can provide theoretical basis for different MRI 3D scanning scenarios.
	
	\item Experiments with acquired 3T and 7T MRI data show that, with the same sampling rates, the quantitative and qualitative results of the recovered MR images by our 1D probabilistic undersampling pattern noticeably outperform those by  existing state-of-the-art undersampling strategies.
\end{itemize}

\section{Method} \label{sec:proposed_method}

In this study, the proposed cross-domain framework includes three parts: the 1D probabilistic undersampling layer, the 1D inverse Fourier transform layer, and the reconstruction network. 

\subsection{1D Probabilistic Undersampling Layer} \label{sec:prob_undersample_layer}

%\begin{figure}[th]
%	\centering
%	\includegraphics[width=0.47\textwidth]{probability_undersample.pdf}
%	\caption{Proposed 1D probabilistic undersampling layer: $\bm{P}_{\rm u}$ is the probability matrix; $\bm{M}_{\rm u}$ is the sampling matrix with binary values \{0,\,1\} only}
%	\label{fig:probability_undersample}
%\end{figure}

\begin{figure*}[th]
	\centering
	\includegraphics[scale=0.6]{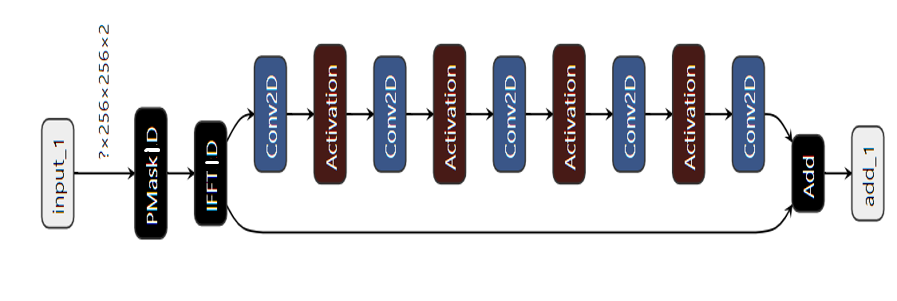}
	\caption{Overall structure of our proposed cross-domain network: the 1D probabilistic undersampling layer, the inverse Fourier transform layer, and the reconstruction network. Undersampling loss ($L_\text{IFT}$) and reconstruction loss ($L_\text{rec}$) are combined for training.}
	\label{fig:undersample_vdsr_network}
\end{figure*}

To simulate the process of $ k $-space undersampling in real scenarios, we propose the 1D probabilistic undersampling layer, as shown in Fig.~\ref{fig:undersample_vdsr_network}, where the input are the fully-sampled $ k $-space data (real and imaginary parts). After undersampling, the output of this layer are the undersampled $ k $-space data, then it can be passed to the next layer. To keep the data format and dimension consistent in training, we separate the real and imaginary parts as two matrices, then concatenate them as two channels: 
\begin{equation}
\fmat{X}_{\rm in} = \big[ \mathrm{real}(\mathcal{K}), \ \mathrm{imag}(\mathcal{K}) \big] \Rdim{m \times n \times 2}.
\end{equation}
This strategy \cite{Schlemper2017Cascade, Han2018kspace, Quan2018Compressed} can effectively avoid the computation of complex numbers.

In this study, the undersampling process is realized by the Hadamard product, which is respectively applied to the real and imaginary parts, as shown in Fig.~\ref{fig:undersample_vdsr_network} and Eq.~\eqref{eq:hardamard_undersample}:
\begin{equation}
\begin{aligned}
	\fmat{X}_{\rm u} &= \fmat{X}_{\rm in} \circ \bm{M}_{\rm u} \\
	&= \big[ \mathrm{real}(\mathcal{K}) \circ \bm{M}_{\rm u}, \ \mathrm{imag}(\mathcal{K}) \circ \bm{M}_{\rm u} \big] \Rdim{m \times n \times 2},
\end{aligned}
  \label{eq:hardamard_undersample}
\end{equation}
where ``$\circ$'' denotes the Hadamard product and $\bm{M}_{\rm u} \Rdim{m \times n}$ denotes the sampling matrix, which has the same dimension as $ k $-space data and ranges from binary values $ \{0, 1\} $. Here, ``0'' denotes the unused part, shown as the black area in the ``mask'' of Fig.~\ref{fig:undersample_vdsr_network}; ``1'' denotes the sampled part, shown as the white dots in the ``mask'' of Fig.~\ref{fig:undersample_vdsr_network}, representing the set of all sampled points in the undersampling pattern.
The effect of the Hadamard product is equivalent to preserving partial $ k $-space data by the sampling matrix $\bm{M}_{\rm u}$ and discarding other parts, then filled with zeros. To this end, the dimension of output is the same as its input, but with only limited rate of $ k $-space data used. The sampling rate is defined as the number of sampling points in $\bm{M}_{\rm u}$ divided by the total number of elements in $\bm{M}_{\rm u}$. 

Instead of directly using existing undersampling patterns, we propose a probability matrix, denoted as $\bm{P}_{\rm u} \Rdim{m \times n}$, ranging from the continuous interval $ [0,1] $. By applying the Bernoulli distribution to each element of $ \bm{P}_{\rm u} $, we obtain
\begin{equation}
\mathcal{B}(1,p=\bm{P}_{\rm u} (x,y)) = 
\begin{cases}
\mathcal{B}(\bm{M}_{\rm u} (x,y) = 1) = p, \\
\mathcal{B}(\bm{M}_{\rm u} (x,y) = 0) = 1 - p. \\
\end{cases} 
\end{equation}
In this manner, we update each element of $ \bm{M}_{\rm u} $ by $ \bm{P}_{\rm u} $. Every $ \bm{M}_{\rm u} (x,y)$ follows the Bernoulli distribution with probability $\bm{P}_{\rm u} (x,y)$, and they are all independently distributed. Similarly, $\bm{P}_{\rm u} (x,y)$ can be considered as the indicator of the importance at the point $ \bm{M}_{\rm u} (x,y) $. 
Thus, only $\bm{P}_{\rm u}$ is trainable in our 1D probabilistic undersampling layer. Because the values in $ \bm{M}_{\rm u} $ are binary (either ``0'' or ``1'') in forward pass, they are natively non-differentiable and have no gradient in backpropagation. Section~\ref{sec:solving_non_differentiable_undersampling} will provide our solution to this issue.

\subsection{Inverse Fourier Transform Layer} \label{sec:inverse_Fourier_transform_layer}

To connect the Fourier domain with the image domain, we design the 1D inverse Fourier transform layer. Since the input of this layer are the undersampled $ k $-space data, we obtain the undersampled MR images via the inverse Fourier transform. The forward and inverse process can be expressed as: 
\begin{align}
\fmat{X}_{\rm u}(u, v) & = \sum_{x=0}^{m-1} \sum_{y=0}^{n-1} \bm{X}_{\rm u}(x, y) \mathrm{e}^{{-i 2 \pi\left(\frac{u x}{m}+\frac{v y}{n}\right)}} , \label{eq:forward_Fourier_transform} \\
\bm{X}_{\rm u}(x, y) & = \frac{1}{m n} \sum_{u=0}^{m-1} \sum_{v=0}^{n-1} \fmat{X}_{\rm u}(u, v) \mathrm{e}^{{i 2 \pi \left(\frac{u x}{m}+\frac{v y}{n}\right)}}. \label{eq:inverse_Fourier_transform} 
\end{align}
Note that Eqs.~\eqref{eq:forward_Fourier_transform} and \eqref{eq:inverse_Fourier_transform} are dual, except for an extra normalized coefficient $\frac{1}{m n}$ in the inverse transform. It can be seen that every sampling point in $ k $-space contains the global information of an MR image in the image domain. Likewise, each pixel in an MR image is generated by using all $ k $-space data. Eqs.~\eqref{eq:forward_Fourier_transform} and \eqref{eq:inverse_Fourier_transform} can be simplified in the matrix form:
\begin{align}
\bm{F}_n &= \begin{bmatrix}
1      & 1            & 1               & \cdots & 1                   \\
1      & \omega       & \omega^{2}      & \cdots & \omega^{n-1}        \\
\vdots & \vdots       & \vdots          & \ddots & \vdots              \\
1      & \omega^{n-1} & \omega^{2(n-1)} & \cdots & \omega^{(n-1)(n-1)}
\end{bmatrix}, \\
 \omega &= \exp \left ( {-i 2 {\pi} / {n}} \right ) , \\
\fmat{X}_{\rm u} &= \bm{F}_m \bm{X}_{\rm u} \bm{F}_n , \\
\bm{X}_{\rm u} & = \bm{F}_m^{-1} \fmat{X}_{\rm u} \bm{F}_n^{-1} = \frac{1}{m n} \bm{F}_m^{\rm H} \fmat{X}_{\rm u} \bm{F}_n^{\rm H} ,
\end{align}
where $\bm{F}_n$ is the Fourier matrix and $ (\cdot)^{\rm H} $ denotes the Hermitian transpose. Since the dimension of the Fourier matrix can be preset, $ \bm{F}_m $ and $ \bm{F}_n $ are computed in advance and not involved in training. Thus, they are considered as constants and are not required to update. The gradient with respect to the input of the inverse Fourier transform can be expressed as
\begin{equation}
\frac{\partial }{\partial \fmat{X}_{\rm u}} = \frac{\partial }{\partial \bm{X}_{\rm u}} \frac{\partial \bm{X}_{\rm u}}{\partial \fmat{X}_{\rm u}} = \frac{1}{m n} \bm{F}_m^{\rm H} \frac{\partial }{\partial \bm{X}_{\rm u}} \bm{F}_n^{\rm H}.
\end{equation}
Hence, the proposed 1D inverse Fourier transform layer adds a little computational cost. Since we have separated the real and imaginary parts in $ k $-space, we can avoid the complex numbers by saving the real and imaginary components of $ \bm{F}_m $ and $ \bm{F}_n $ respectively to accelerate the training period. 
%According to the Euler formula, we obtain
%\begin{align}
%\bm{F}_n^{\cos} &= \begin{bmatrix}
%1      & 1            & 1               & \cdots & 1                   \\
%1      & \eta       & \eta^{2}      & \cdots & \eta^{n-1}        \\
%\vdots & \vdots       & \vdots          & \ddots & \vdots              \\
%1      & \eta^{n-1} & \eta^{2(n-1)} & \cdots & \eta^{(n-1)(n-1)}
%\end{bmatrix},  \label{eq:fourier_matrix_cos} \\
%\bm{F}_n^{\sin} &= \begin{bmatrix}
%1      & 1            & 1               & \cdots & 1                   \\
%1      & \kappa       & \kappa^{2}      & \cdots & \kappa^{n-1}        \\
%\vdots & \vdots       & \vdots          & \ddots & \vdots              \\
%1      & \kappa^{n-1} & \kappa^{2(n-1)} & \cdots & \kappa^{(n-1)(n-1)}
%\end{bmatrix}, \label{eq:fourier_matrix_sin} \\
% \eta &= \cos  ( { 2 {\pi}{n}}  ) , \quad  \kappa= - \sin  ( { 2 {\pi}{n}}  ) .
%\end{align}

\subsection{Reconstruction Network} \label{sec:reconstruction_network}

Fig.~\ref{fig:undersample_vdsr_network} illustrates the structure of our proposed reconstruction network (RecNet). It is placed after the inverse Fourier transform layer. In this study, we adopt a common CNN structure with a global skip connection:
\begin{equation}
\bm{X}_{\rm rec} = \bm{X}_{\rm u} + f_\mathrm{cnn} (\bm{X}_{\rm u} | \bm{\theta}),
\end{equation}
where the depth of CNN is 10, Section~\ref{sec:effect_reconstruction_network} explains how we choose the depth. Except the last convolutional layer (kernel 1$\times$1, channel  = 1, stride = 1, pad = 0)  for feature fusion, other convolutional layers have the setting of kernel 3$\times$3, channel = 16, stride = 1, and pad = 0, followed by the rectified linear unit (ReLU) activation to keep nonlinear capability. The residual learning scheme is adopted to alleviate the gradient vanishing problem, with more details provided in \cite{Kim2016VDSR}. After training, the output of RecNet $ \bm{X}_{\rm rec} $ can approximate the ground-truth MR image $ \bm{Y}_{\rm rec} $. To obtain the universal results of sampling matrices, we adopt the conventional Euclidean loss to train our model, and this loss is defined as the reconstruction loss: 
\begin{equation}
L_\text{rec} = \frac{1}{2} \| \bm{X}_{\rm rec} - \bm{Y}_{\rm rec} \|_\fro^2,
\end{equation}
where $\| \cdot \|_\fro$ denotes the Frobenius norm. 
Unlike other deep learning based MRI reconstruction methods \cite{Wang2016Accelerating, Yang2018DAGAN, Mardani2017Deep, Hyun2018Deep, Quan2018Compressed, Schlemper2017Cascade, Zhang2017ISTA, pereira2016brain}, 
%(Fig.~\ref{fig:DL_based_MRI_reconstruction_methods}), 
we focus on discovering and analyzing the undersampling pattern, rather than designing complicated CNN structures. 

\subsection{Stable Constraints for Undersampling Patterns} \label{sec:stable_constraint}

\begin{figure}[th]
	\centering
%	\subfloat[10\% sampling matrix with Eq.~\eqref{eq:mean_sample_rate_constraint} only (initial status)]{\includegraphics[width=0.40\linewidth]{random_sample_mask.pdf} 	
%		\label{subfig:random_sample_mask}} \hfil
%	\subfloat[10\% sampling matrix with Eqs.~\eqref{eq:mean_sample_rate_constraint}--\eqref{eq:distance_constraint} (initial status)]{\includegraphics[width=0.40\linewidth]{uniform_sample_mask.pdf}
%		\label{subfig:uniform_sample_mask}}
	\subfloat[10\% sampling matrix with Eq.~\eqref{eq:mean_sample_rate_constraint} only (initial status)]{\includegraphics[width=0.45\linewidth]{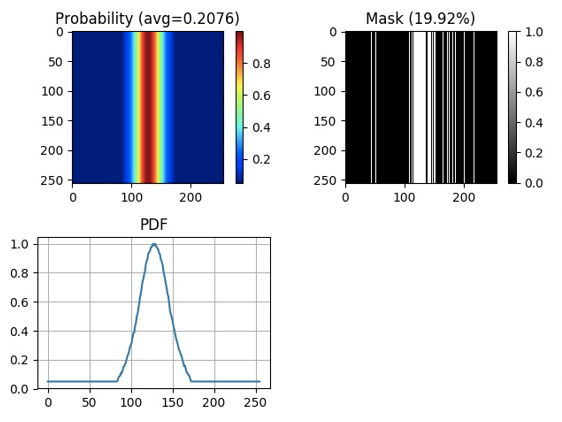} 	
		\label{subfig:random_sample_mask}} \hfil
	\subfloat[20\% sampling matrix with Eqs.~\eqref{eq:mean_sample_rate_constraint}--\eqref{eq:distance_constraint} (initial status)]{\includegraphics[width=0.45\linewidth]{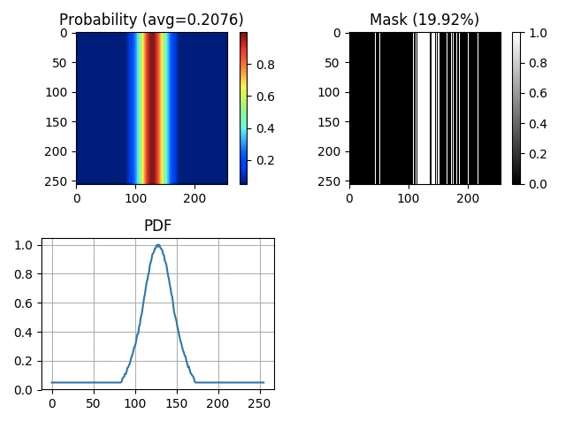}
		\label{subfig:uniform_sample_mask}}
	\caption{An example of 10\% sampling matrices with/out  constraints}
	\label{fig:sample_mask_constraint}
\end{figure}

%We use probability matrix $ \bm{P}_{\rm u} $ to generate sampling matrix $ \bm{M}_{\rm u} $ by the Bernoulli distribution, to introduce significant randomness. This helps $ \bm{P}_{\rm u} $ to be fully trained. 
After training, we require stable sampling matrix $ \bm{M}_{\rm u} $ generated by probability matrix $ \bm{P}_{\rm u} $ to analyze its pattern. However, randomness from the Bernoulli distribution will interfere at this stage. As shown in Fig.~\ref{subfig:random_sample_mask}, sampling matrix $ \bm{M}_{\rm u} $ is generated by probability matrix $ \bm{P}_{\rm u} = 10\%$ with Bernoulli distribution. It can be seen that those sampling points are randomly placed, where some areas are sparse but others are dense. This makes  sampling matrix $ \bm{M}_{\rm u} $ unstable and difficult to reproduce. Every $ \bm{M}_{\rm u} $ generated by the same $ \bm{P}_{\rm u} $ is quite different from each other, so that the performance of the RecNet varies dramatically. This apparently prevents probability matrix $ \bm{P}_{\rm u} $ from convergence. 

%According to the definitions in physics, the distance between gas molecules changes drastically, similar to Fig.~\ref{subfig:random_sample_mask}; however, the distance between solid molecules is relatively fixed, similar to Fig.~\ref{subfig:uniform_sample_mask}. 
To overcome the interference caused by randomness, we propose the following stable constraints and they are applied in the stage when probability matrix $ \bm{P}_{\rm u} $ generates sampling matrix $ \bm{M}_{\rm u} $ ($\bm{P}_{\rm u} \rightarrow \bm{M}_{\rm u}$): 

	1)
	\textbf{Total sampling rate constraint}. During training, the average of the values in probability matrix $ \bm{P}_{\rm u} $ should be equal to the preset total sampling rate: 
%	Since all experiments in this paper are based on the fixed sampling rate, Eq.~\eqref{eq:mean_sample_rate_constraint} is considered as an essential constraint. 
	\begin{equation}
	\left \| \bar{\bm{P}_{\rm u}} - \text{rate} \right \|_2 < \epsilon,  \label{eq:mean_sample_rate_constraint}
	\end{equation}
	where $ \bar{\bm{P}_{\rm u}} $ is the average of the probability matrix, $ \epsilon = 0.1\%$ is the tolerance and assures that the difference between the actual sampling rate and the preset value is smaller than 0.1\%.
	
	2)
	\textbf{Regional sampling distance constraint}. In each region $\mathcal{C} \Rdim{10 \times 10}$, the number of sampling points is allocated according to its probability, i.e., the larger the probability is, the more the sampling points will be. The distances between those sampling points with the same probability should be nearly uniform, ranging from $[r_0, 2 r_0]$. Eqs.~\eqref{eq:probability_constraint} and \eqref{eq:distance_constraint} can apparently influence the final pattern in  $ \bm{M}_{\rm u} $.  
	\begin{gather}
	\left \| \bar{p} - \frac{ \sum_{x,y \in \mathcal{C}} | \bm{M}_{\rm u} (x, y) |}{m n}  \right \|_2 < \epsilon,   \label{eq:probability_constraint} \\ 
	r_0 < \left \| \bm{M}_{\rm u} (x_i, y_j) - \bm{M}_{\rm u} (x_k, y_l) \right \|_2 < 2 r_0, \label{eq:distance_constraint} \\
	\forall \, i, k = 1, 2, \ldots, m, \quad \forall \, j, l = 1, 2, \ldots, n. 
	\end{gather}
	where $ \bar{p} $ denotes the actual mean value of region $ \mathcal{C} \Rdim{10 \times 10} $ and $ r_0 $ denotes the preset minimal distance of all sampling points. According to Appendix~\ref{apdx:curve_fitting_p_r0}, we obtain $ \bar{p} = \frac{\sqrt{2}}{10} r_0^2 - \frac{\sqrt{2}}{2} r_0 + 1.0$.

Since the actual sampling distance is integer, distributing sampling points in $ \bm{M}_{\rm u} $ will bring in some localization error,  but we confirm this error is negligible by rigorously assuring that $ \| \bar{\bm{P}_{\rm u}} - \text{rate} \|_2 < \epsilon $ satisfies. 

%In summary, our stable constraints include: 1) total sampling rate constraint~\eqref{eq:mean_sample_rate_constraint}; 2) regional sampling distance constraint~\eqref{eq:probability_constraint} and \eqref{eq:distance_constraint}. 
%They are applied in the stage when probability matrix $ \bm{P}_{\rm u} $ generates sampling matrix $ \bm{M}_{\rm u} $. As shown in Fig.~\ref{fig:probabilistic_undersample_layer_compute}, this stage is not involved in the forward pass and the backpropagation. 
%%Thus, when training, the processes of $\bm{P}_{\rm u} \rightarrow \bm{M}_{\rm u}$, forward pass, and backpropagation are alternately executed. 
%As shown in Fig.~\ref{subfig:uniform_sample_mask}, with constraints~\eqref{eq:mean_sample_rate_constraint}--\eqref{eq:distance_constraint}, sampling matrix $ \bm{M}_{\rm u} $ is generated by probability matrix $ \bm{P}_{\rm u} = 10\%$. It reveals that all sampling points in $ \bm{M}_{\rm u} $ are uniformly distributed, where no sparse or dense area exists. Furthermore, with the stable constraints, the sampling matrix is quite easy to reproduce. 
%%Section~\ref{sec:effect_stable_constraint} will demonstrate the effect of these stable constraints.

\section{Experiments} \label{sec:experiments}

\begin{figure}[th]
	\centering
	\subfloat[J-CUR \cite{Weiss2020Joint}]{\includegraphics[width=0.28\linewidth]{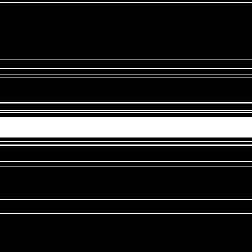} 
		\label{subfig:J_CUR_mask}} \hfil
	\subfloat[PFGMS \cite{Gozcu2018LearningMRI}]{\includegraphics[width=0.28\linewidth]{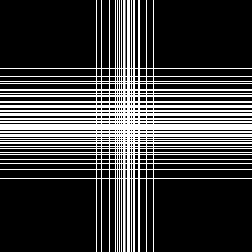}
		\label{subfig:PFGMS_mask}} \hfil
	\subfloat[J-MoDL \cite{Aggarwal2019J_MoDL}]{\includegraphics[width=0.28\linewidth]{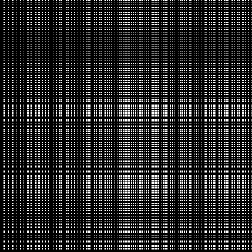}
		\label{subfig:JMoDL_mask}} \\
%	\subfloat[LOUP \cite{Bahadir2019Learning}]{\includegraphics[width=0.28\linewidth]{LOUPE_mask.pdf}
%		\label{subfig:compared_LOUP_mask}} \hfil
%	\subfloat[Gaussian  \cite{Cook1986Gaussian}]{\includegraphics[width=0.28\linewidth]{gaussian_mask.pdf}
%		\label{subfig:compared_gaussian_mask}} \hfil
%	\subfloat[Poisson  \cite{Jones2006Poisson}]{\includegraphics[width=0.28\linewidth]{poisson_mask.pdf}
%		\label{subfig:compared_poisson_mask}} 
	\caption{Several existing $ k $-space undersampling patterns used in this study for comparison}
	\label{fig:comapred_undersampled_mask}
\end{figure}

Our source code is availably online\footnote{https://github.com/xueshengke/1D-PUPO}. It is based on Tensorflow \cite{Abadi2016Tensorflow} with Keras APIs. Our experiments were executed on a Ubuntu Linux server equipped with an Intel Xeon(R) Platinum CPU @ 2.50~GHz, total 528~GB memory, and four NVIDIA Tesla V100 (32~GB) GPUs. Our 1D probabilistic undersampling strategy was compared with several existing state-of-the-art undersampling methods with sampling rates 10\%--50\%. The compared undersampling methods are J-CUR \cite{Weiss2020Joint} (Fig.~\ref{subfig:J_CUR_mask}), PFGMS \cite{Gozcu2018LearningMRI} (Fig.~\ref{subfig:PFGMS_mask}), J-MoDL \cite{Aggarwal2019J_MoDL} (Fig.~\ref{subfig:JMoDL_mask}), LOUP \cite{Bahadir2019Learning} (Fig.~\ref{subfig:compared_LOUP_mask}), Gaussian sampling \cite{Cook1986Gaussian} (Fig.~\ref{subfig:compared_gaussian_mask}), and Poisson sampling \cite{Jones2006Poisson} (Fig.~\ref{subfig:compared_poisson_mask}).
For fair comparison, we replaced our probability and sampling matrices with those compared (either learned or fixed) patterns, and then trained our cross-domain model. As the model parameters trained in different undersampling cases could not be shared, we trained multiple models individually for various sampling rates. 

In addition, we aimed to optimize the undersampling pattern and to discover the optimal probability distribution through our 1D probabilistic undersampling layer, rather than to design complicated CNN structures. Consequently, we did not conduct the architecture search for the RecNet. 
%Section~\ref{sec:effect_stable_constraint} will prove that the quality of recovered MR images is improved mainly by the learned 1D probabilistic undersampling pattern, secondly by the trained RecNet. 

\subsection{Training Configurations and Initialization} \label{sec:configuration_initialization}

At the initial stage of the 1D probabilistic undersampling layer, all values in probability matrix $ \bm{P}_{\rm u}$ were equal to the target sampling rate, i.e., $ \bm{P}_{\rm u} = $ 10\% or 20\%. Eq.~\eqref{eq:mean_sample_rate_constraint} (i.e., $ \bar{\bm{P}_{\rm u}} \approx \text{rate} $) was guaranteed during training. The initial status of sampling matrix $ \bm{M}_{\rm u} $ was decided by probability matrix $ \bm{P}_{\rm u} $ with Bernoulli distribution. 
%As the training progresses, $ \bm{M}_{\rm u} $ will be stable when $ \bm{P}_{\rm u} $ converges. 
Since each value of probability matrix $ \bm{P}_{\rm u} $ followed $ p \in [0,1]$, we proposed an upper bound $ p \leq P_{\max} = 1 $, in case that the probability overflowed. In addition, the Bernoulli distribution made no sense when $p = 0$, since the corresponding values in sampling matrix $ \bm{M}_{\rm u} $ could not be updated. This lead to a local minimum and poor performance of the RecNet. 
To this end, we proposed a lower bound $ p \geq  P_{\min}$. Together with the total sampling rate constraint~\eqref{eq:mean_sample_rate_constraint}, Section~\ref{sec:analysis_optimal_undersampling_pattern} will demonstrate how to decide the precise value of $ P_{\min} $. 

%Since no trainable parameter exists in the inverse Fourier transform layer, we only need to preserve the Fourier matrices of Eqs.~\eqref{eq:fourier_matrix_cos} and \eqref{eq:fourier_matrix_sin} during initialization. 

Our cross-domain model was trained  by using $ L_\text{joint} $ ($ \lambda_1 = \lambda_2 = 1 $) to analyze the undersampling patterns and compare the quality of final MR images. 

%Our cross-domain model is trained  in two conditions: 1) using $ L_\text{IFT} $ ($ \lambda_1 = 1, \lambda_2 = 0 $); 2) using $ L_\text{joint} $ ($ \lambda_1 = \lambda_2 = 1 $) to analyze the undersampling patterns and compare the quality of final MR images. 

The training configurations of our model were summarized in Table~\ref{tab:MR_rec_train_config}. 
%We set the initial learning rate as $10^{-3}$, which will be divided by  $\sqrt{10}$ in every 20 epochs, until it reaches the minimum $10^{-8}$. We use the Adam \cite{Kingma2014Adam} optimizer with default coefficients $\beta_1 = 0.9$ and $\beta_2 = 0.999$. In addition, we adopt the early-stopping strategy: the training will terminate in advance if no improvement occurs in 20 successive epochs. 
Some settings of the RecNet have been given in Section~\ref{sec:reconstruction_network}. 
%To keep the dimensions of input and output identical, we adopt the zero-padding scheme in each convolutional layer. 
We set the depth of the RecNet as 10, denoted as RecNet-10. Section~\ref{sec:effect_reconstruction_network} will demonstrate how we select the depth of the RecNet.

\begin{table}[ht] \footnotesize 
	\centering
	\caption{Training configurations of our cross-domain model} \label{tab:MR_rec_train_config}
	\addtolength{\tabcolsep}{-1pt}
	\begin{tabular}{lc||lc}
		\toprule
		Description           &    Value    & Description           &           Value            \\ \midrule
		Batch                 &     16      & Optimizer             & Adam \cite{Kingma2014Adam} \\
		Maximal epoch         &     200     & $\beta_1$ of Adam     &            0.9             \\
		Initial learning rate &   1e$-$3    & $\beta_2$ of Adam     &           0.999            \\
		Learning policy       &    Step     & Minimal learning rate &           1e$-$8           \\
		Learning decay        & $\sqrt{10}$ & Weight decay          &           1e$-$5           \\
		Decay step            &     20      & Device                &            GPU             \\ \bottomrule
	\end{tabular} 
\end{table}

\subsection{Datasets} \label{sec:datasets}

We adopted the 3T MR images from Multi-modal Brain Tumor Segmentation Challenge 2018 \cite{Liu2016Periacetabular, Bakas2017Advancing}, as the training and validation sets, denoted as BraTS 2018. All image slices were obtained by 3D scanning. Specifically, it contains 200 training volumes and 50 validation volumes. Each volume comes from one patient/candidate. The spatial resolution is 256$\times$256$\times$(240--320), the minimal voxel is 1\,mm$^3$. We chose 200 volumes for training and 10 out of 50 volumes for validation. 
%Fig.~\ref{fig:BraTS_example} shows an example of BraTS 2018. 
The pixels' intensity was normalized to $[0, 1]$. 

%\begin{figure}[th]
%	\centering
%	\subfloat[Axial]{\includegraphics[height=0.27\linewidth]{BraTS_Axial.png}} \hfil
%	\subfloat[Coronal]{\includegraphics[height=0.27\linewidth]{BraTS_Coronal.png}} \hfil
%	\subfloat[Sagittal]{\includegraphics[height=0.27\linewidth]{BraTS_Sagittal.png}} 	
%	\caption{An example of brain MR images from BraTS 2018 \cite{Liu2016Periacetabular, Bakas2017Advancing}}
%	\label{fig:BraTS_example}
%\end{figure}

As for the test set,  we acquired 3D T1-weighted MRI data  (brain only) from healthy volunteers on both 3T (SIEMENS Prisma) and 7T (SIEMENS Magnetom) scanners, denoted as Test3T\&7T. Here both the fully-sampled k-space data (Cartesian sampling) and MR images were obtained and kept. 
It consists of 20 volumes of $ k $-space data from 10 different volunteers. Each one was scanned in both 3T and 7T MRI. The spatial resolutions are 256$\times$256$\times$192 for 3T and 310$\times$314$\times$208 for 7T, respectively. We used MRI data from two-field MRI scanners with different resolutions to prove the generalization of our proposed method. 

Since we focused on the human tissue and ignored the background in MR images, we proposed the following data argumentation strategy:
\begin{enumerate}
	\item \textbf{Translation}. All contents of tissue were translated to the geometrical center of MR images.
	\item \textbf{Rotation}. All images were randomly rotated [0, 360) degrees by 8 times. 
\end{enumerate}
This strategy was used to prevent the over-fitting problem and boosted the generalization capability of our model. Because most of images in BraTS 2018 are similar and have fixed shapes, 
%except for small difference from various samples, 
the diversity of $ k $-space data is obviously insufficient. To obtain the universal probabilistic undersampling patterns after training, our data argumentation strategy (translation and random rotation) significantly enriches the diversity of our training set. Section~\ref{sec:analysis_optimal_undersampling_pattern} shows that our probability matrices trained with various sampling rates are perfectly symmetric. 
%BraTS2018 数据集的图像中组织的形态、位置均是非常接近的，所以对应的$ k $空间中能量分布的多样性不足。为了在实验中通过训练得到概率化降采样轨迹的通用规律，本章采用的数据扩充策略，尤其是平移和随机角度 (0\textdegree--360\textdegree) 旋转，显著弥补了数据量过少而导致的多样性缺失的弊端。在第~\ref{sec:analyze_probability_mask} 节中，由不同采样比例的条件下训练得到的概率矩阵的可视化结果充分体现了概率矩阵的完美对称性和通用性。

To keep the dimension of training data consistent, we resized  all MR images to 256$\times$256  and concatenated them along the sagittal direction. Thus, during training and validation, the dimension of one batch became 256$\times$256$\times N$, where $N$ is the batch size. 

Since the $ k $-space data were used as the input of our model while BraTS 2018 only provided MR images, we converted all training and validation images into the frequency domain via the 1D fast Fourier transform. As for the test set, we directly used 3T/7T fully-sampled $ k $-space data. To avoid complex values, we split the real and imaginary parts, then combined them as two channels. The MR images reconstructed from fully-sampled $ k $-space data in training and test sets were used as the ground-truth. 

%Fig.~\ref{fig:example_MR_Fourier_image} shows an example of the sagittal brain MR image and its Fourier spectrum. 

%由于本章模型的输入是 $k$ 空间数据，而 BraTS 2018 图像集只提供已经重建的MR图像，本章首先将所有的训练和验证图像通过 1D 的 FFT 函数转换至 Fourier 空间。为了避免使用复数格式的数据，本章将它们的实部和虚部分离，并组合为两个通道，以模拟 $k$ 空间数据的计算方式。对于测试图像集，本章使用 3T 和 7T、T1 和 T2 双模态、全采样3D扫描的 $k$ 空间数据，把实部和虚部分离并重新组合后，作为本章模型的输入。图~\ref{fig:example_MR_Fourier_image} 展示了头部 MR 矢状面图像和其 Fourier 频谱的一个示例。其中，中心低频部分的能量明显大于边缘高频部分的能量。可以发现，由于边缘突变效应，图~\ref{subfig:MR_Fourier_real_part} 和图~\ref{subfig:MR_Fourier_image_part} 中都存在一条经过频谱中心的竖线，这表明了$ k $空间中需要过多的频率分量近似MR图像的底部边缘组织。此情况不利于降采样轨迹的分析，也影响了数据扩充策略的效果 (部分组织因旋转超出边缘而丢失)，会导致训练获得的概率矩阵和采样矩阵有偏差。所以，本章通过将图像组织部分平移至中心的预处理方式来解决这个问题。由于本章模型的目标是重建图像，BraTS 2018 训练、验证集和 Test3T\&7T 测试集中的 MR 图像自身即可作为标签来使用。

%\begin{figure}[th]
%	\centering
%	\subfloat[Gray image]{\includegraphics[width=0.37\linewidth]{MR_image_gray} 	
%		\label{subfig:MR_image_gray}} \hfil
%	\subfloat[Color image]{\includegraphics[width=0.37\linewidth]{MR_image_color}
%		\label{subfig:MR_image_color}} \\ [-1mm]
%	\subfloat[Real part]{\includegraphics[width=0.37\linewidth]{MR_image_fourier_real}
%		\label{subfig:MR_Fourier_real_part}} \hfil
%	\subfloat[Imaginary part]{\includegraphics[width=0.37\linewidth]{MR_image_fourier_image}  	
%		\label{subfig:MR_Fourier_image_part}}
%	\caption{An example of the sagittal brain MR image and its Fourier spectrum}
%	\label{fig:example_MR_Fourier_image}
%\end{figure}

\subsection{Evaluation Metric} \label{sec:metric}

Peak signal-to-noise ratio (PSNR) has been commonly used to measure the quality of MR images in various computer vision tasks. PSNR is defined as follows:
\begin{align}
{\rm MSE}  &= \frac{1}{N} \| \bm{I}_{\rm rec} - \bm{I}_{\rm gnd}\|_\fro^2, \\
{\rm PSNR} &= 10 \cdot \log_{10} \left( \frac{P_{\max}^2}{{\rm MSE}} \right) \ {\rm dB}, \label{eq:PSNR}
\end{align}
where $\bm{I}_{\rm rec}$ is the recovered image, $\bm{I}_{\rm gnd}$ is the original image, and $P_{\max}$ is the maximum pixel value (normally 1.0) of an image. A high-quality MR image usually has a large value of PSNR and a lower value of mean-squared error (MSE). Although structural similarity index (SSIM) \cite{Wang2004SSIM} is often adopted to evaluate the quality of MR images, we found that the most values of SSIM of recovered MR images were nearly equal to 1.0, which were too difficult to distinguish. Thus,  we used only PSNR as the effective metric to evaluate the performances of different methods. 

In our model, we adopted two loss functions: undersampling loss $ L_\text{IFT} $ and reconstruction loss $ L_\text{rec} $, for supervised training. Correspondingly, we presented two indicators: undersampling PSNR ($ \bm{X}_{\rm u}$ vs.\ $ \bm{Y}_{\rm rec} $) and reconstruction PSNR ($ \bm{X}_{\rm rec}$ vs.\ $ \bm{Y}_{\rm rec} $). 

%SSIM 的具体定义如下：
%\begin{align}
%{\rm SSIM} &= \frac{1}{MN} \sum_{m} \sum_{n} \frac{(2\mu_m\mu_n +　\rho_1)(2\sigma_{mn}+\rho_2)}{(\mu_m^2 + \mu_n^2 + \rho_1)(\sigma_m^2 + \sigma_n^2 + \rho_2)}, \label{eq:SSIM} \\
%& m = 1,2,\ldots,M, \quad n = 1,2,\ldots,N,
%\end{align}
%其中 $\mu_m$ 和 $\mu_n$ 分别表示 $m$ 和 $n$ 的均值，$\sigma_m^2$ 和 $\sigma_n^2$ 分别是 $m$ 和 $n$ 的方差，$\sigma_{mn}$ 是 $m$, $n$ 的协方差， $\rho_1 = (\kappa_1 R)^2$ 和 $\rho_2 = (\kappa_2 R)^2$ 都是维持稳定的常量，$R$ 表示像素的动态范围，$\kappa_1 = 0.01$，$\kappa_2 = 0.03$。SSIM 的取值范围是 [0, 1]，其值越大表示恢复图像的结构性越好，细节越丰富。当两张图像 $\bm{I}_x$ 和 $\bm{I}_y$ 完全一致时，则 ${\rm SSIM}(\bm{I}_x, \bm{I}_y) = 1$。

\section{Results and discussions} \label{sec:results_and_discussions}

\subsection{Effect of Reconstruction Network} \label{sec:effect_reconstruction_network}

Table~\ref{tab:comparison_depth_RecNet} shows the PSNR of our 1D probabilistic undersampling layer with different depths of RecNet and sampling rate 10\%--50\%. These results were obtained from the Test3T\&7T set. It can be seen that the difference between the actual sampling rate and the target value is quite small ($<$0.1\%), which proves the effectiveness of our total sampling rate constraint~\eqref{eq:mean_sample_rate_constraint}. As the sampling rate or the depth of RecNet increases, the undersampling PSNR and reconstruction PSNR both improve significantly. Note that the improvement first comes from our 1D probabilistic undersampling patterns, then from our trained RecNet. The deeper the RecNet is, the better the quality of MR images we can obtain. In addition, when the sampling rates are identical, the undersampling PSNR also improves with the increasing of the depth of RecNet, which indicates that our cross-domain training scheme can simultaneously optimize the 1D probabilistic undersampling layer and the RecNet, and obtain matched sampling matrices and parameters of RecNet. 

%表~\ref{tab:comparison_depth_RecNet} 展示了在降采样比例 10\%--50\% 条件下，1D概率化降采样层结合不同深度的重建网络 (RecNet)，基于 Test3T\&7T 的 MR 图像集测试的结果。可以发现，实际的采样比例和期望值只有微小的误差 (小于 0.1\%)，这证实了总采样比例约束 (公式~\eqref{eq:mean_sample_rate_constraint}) 的有效性。随着降采样比例的增加，降采样PSNR和重建PSNR值都在逐渐上升。因为直观理解$ k $空间的采样信息越多，重建的MR图像质量越高。随着网络深度的增加 (RecNet-0表示无重建网络，只靠降采样过程)，降采样PSNR和重建PSNR都在逐渐提高，这说明了提升MR图像的质量主要依靠本章设计的概率化降采样轨迹，其次是重建网络RecNet对图像细节的恢复，并且网络越深，其恢复能力越强。另外，在同样的采样比例下，降采样PSNR也能随着网络加深而提高，这说明了本章的跨域联合训练机制可以同时优化频域的1D概率化降采样层和空间域的重建网络，最终获得一组配对的降采样轨迹和重建网络的参数。

\begin{table}[t] \scriptsize
	\centering
	\caption{PSNR of our 1D probabilistic undersampling layer with various depths of RecNet and sampling rate 10\%--50\%. Results were obtained from the Test3T\&7T set
%		降采样比例 10\%--50\% 条件下，1D概率化降采样层结合不同深度的重建网络 (RecNet)，基于 Test3T\&7T 的 MR 图像集测试的结果。此处仅报告 PSNR，因为 SSIM 差别不大
	} \label{tab:comparison_depth_RecNet}
	\addtolength{\tabcolsep}{-2.5pt}
	\begin{tabular}{cccccc}
		\toprule
		Sampling              &                  \multicolumn{5}{c}{Undersampling PSNR (dB) / Reconstruction PSNR (dB)}                   \\
		\cmidrule{2-6}   
		rate &  RecNet-0  &   RecNet-5    &   RecNet-10   &   RecNet-15   &       RecNet-20        \\ 
		\midrule
		10.09\%     & 32.80 / -- & 33.03 / 36.43 & 34.25 / 37.34 & 34.99 / 38.08 & \textbf{35.33 / 38.74} \\
		20.01\%     & 34.98 / -- & 35.67 / 38.49 & 36.83 / 39.15 & 36.99 / 39.87 & \textbf{37.68 / 40.09} \\
		29.91\%     & 36.06 / -- & 37.01 / 40.91 & 38.70 / 41.46 & 39.04 / 42.34 & \textbf{39.63 / 42.97} \\
		40.04\%     & 37.14 / -- & 39.49 / 41.99 & 40.99 / 43.87 & 41.51 / 44.13 & \textbf{42.03 / 44.55} \\
		50.00\%     & 39.56 / -- & 41.32 / 44.56 & 42.01 / 45.92 & 42.51 / 46.21 & \textbf{43.36 / 46.36} \\ 
		\bottomrule
	\end{tabular}
	\leftline{\scriptsize Note: -0 for no RecNet; -5, -10, -15, -20 for the depths of the RecNet}
\end{table}

%Fig.~\ref{fig:MR_RecNet_different_depth} shows the number of parameters, Flops, and memory of our 1D probabilistic undersampling layer with different depths of RecNet, based on BraTS 2018 training set. Most parameters are in the probability matrix and sampling matrix of our 1D probabilistic undersampling layer, i.e., $ 256 \times 256 \times 2 = 131,072 $. In the RecNet, only a few parameters exist. However, as the depth of RecNet increases, the number of Flops rises rapidly, because the feature maps with the size of 256$\times$256 require a huge amount of computational cost in convolution. Likewise, the memory consumption is proportional to the depth of RecNet during training. To balance the performance and computational burden of our model, we choose the depth as 10 in the rest of experiments. 

\begin{figure}[th]
	\centering
	\subfloat[\#Param vs.\ depth]{\includegraphics[width=0.46\linewidth]{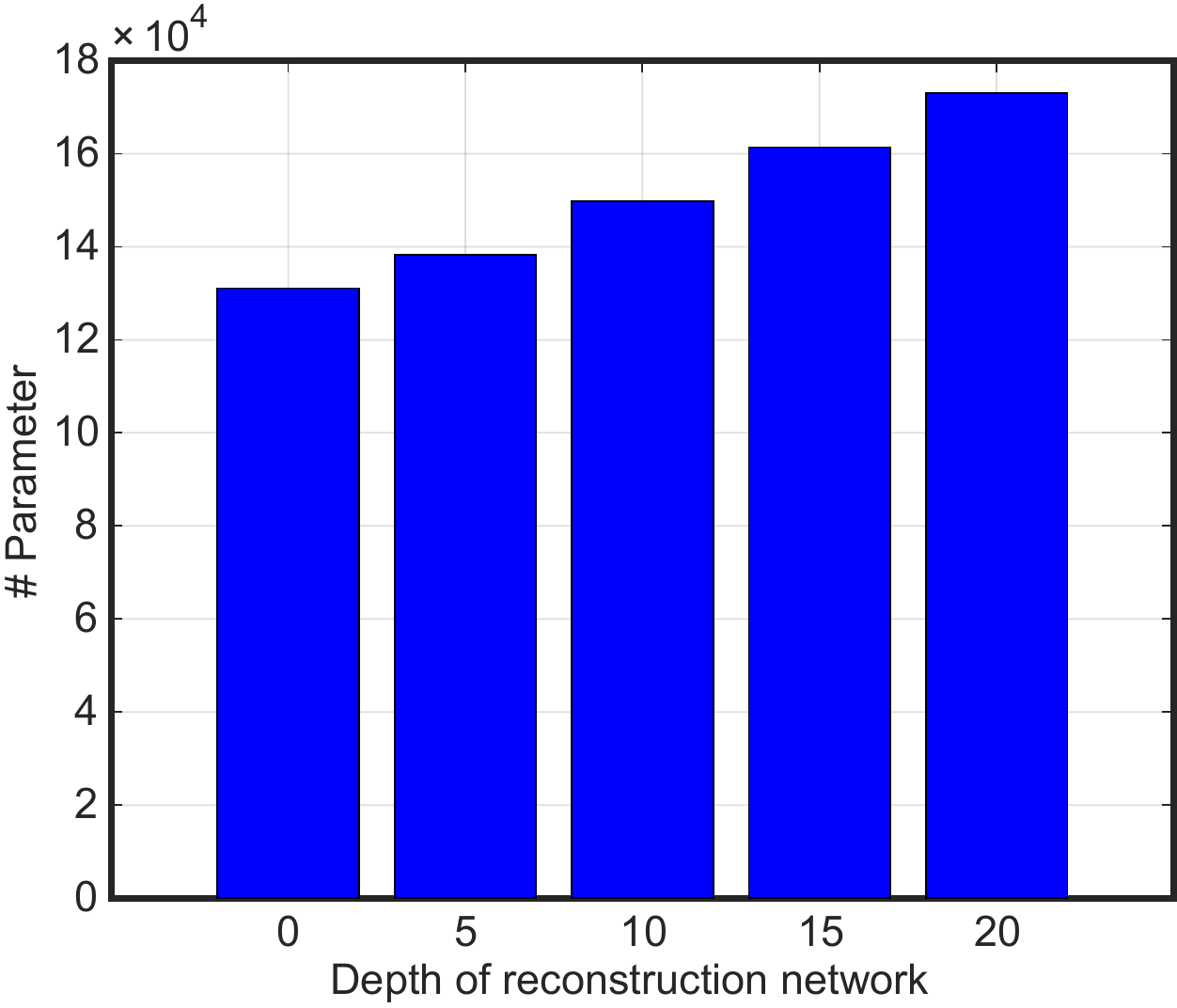} 	
		\label{subfig:MR_RecNet_param_vs_depth}} \hfil
	\subfloat[\#Flops vs.\ depth]{\includegraphics[width=0.46\linewidth]{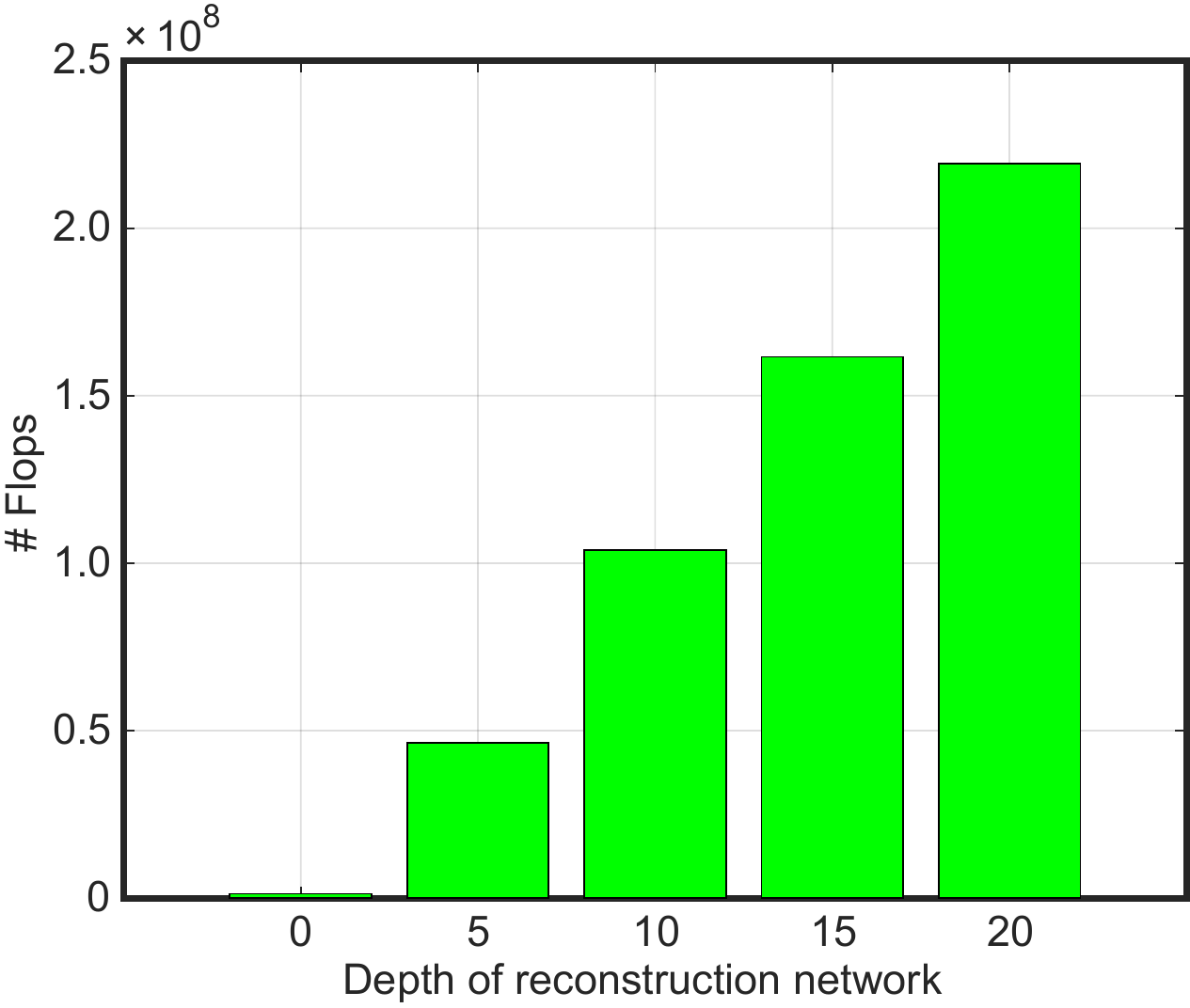}
		\label{subfig:MR_RecNet_flops_vs_depth}} 
%	\subfloat[\#Memory vs depth]{\includegraphics[width=0.46\linewidth]{MR_RecNet_memory_vs_depth.pdf}
%		\label{subfig:MR_RecNet_memory_vs_depth}} 
	\caption{\#Param and \#Flops of our 1D probabilistic undersampling layer with different depths of RecNet obtained from the BraTS 2018 training set
%		基于 BraTS 2018 的 MR 图像集训练, 1D概率化降采样层结合不同深度的重建网络 (RecNet) 模型的指标：参数量、(乘-加) 计算量、和内存占用
}
	\label{fig:MR_RecNet_different_depth}
\end{figure}

Fig.~\ref{fig:MR_RecNet_different_depth} shows the number (\#) of parameters and floating operations per second (flops) of our 1D probabilistic undersampling layer with different depths of the RecNet. These results were based on the BraTS 2018 training set. Most parameters are in the probability matrix and sampling matrix of our 1D probabilistic undersampling layer, i.e., $ 256 \times 256 \times 2 = 131,072 $. In the RecNet, only a few parameters exist. However, as the depth of RecNet increases, the number of flops rises rapidly, because the feature maps with the size of 256$\times$256 require a huge amount of computational cost in convolution. To balance the performance and computational burden of our model, we chose the depth as 10 in the rest of experiments. 

%表~\ref{tab:comparison_parameter_flops_memory_VDSR} 和图~\ref{fig:MR_RecNet_different_depth} 展示了基于 BraTS 2018 的 MR 图像集训练，1D概率化降采样层结合不同深度的重建网络 (RecNet) 的指标：参数量、(乘-加) 计算量、和内存占用。由表~\ref{tab:comparison_parameter_flops_memory_VDSR} 和图~\ref{subfig:MR_RecNet_param_vs_depth} 发现，本章跨域网络的主要参数是 1D 概率化降采样层中的概率矩阵和采样矩阵参数$ 256 \times 256 \times 2 = 131,072 $，而重建网络RecNet中只有相对较少的卷积核和注意力模块的参数。随着重建网络深度的增加，它们并没有引入很多的参数量。然而，由图~\ref{subfig:MR_RecNet_flops_vs_depth} 发现，随着网络深度的增加，本章模型的计算量成倍地迅速增长，这是因为卷积层中的计算量和特征图的维度大小有关，256$\times$256的维度使得卷积过程的计算负担变得很巨大。同理，由图~\ref{subfig:MR_RecNet_memory_vs_depth} 发现，本章模型在训练的过程中占用的内存也与网络的深度呈正比例关系。因此，考虑网络的性能、资源 (时间和内存) 消耗和设备的计算负担，本章最终只采用网络深度为 10 的重建网络 RecNet-10 进行实验比较。

%\begin{table}[th] \small
%	\centering
%	\caption{\#Param, \#Flops, and \#Memory of our 1D probabilistic undersampling layer with different depths of RecNet, based on BraTS 2018 training set
%%		基于 BraTS 2018 的 MR 图像集训练, 1D概率化降采样层结合不同深度的重建网络 (RecNet) 模型的指标：参数量、(乘-加) 计算量、和内存占用
%	}
%	\label{tab:comparison_parameter_flops_memory_VDSR}
%	\addtolength{\tabcolsep}{5pt}
%	\begin{tabular}{cccccc}
%		\toprule
%		Metric     & RecNet-0 &  RecNet-5  &  RecNet-10  &  RecNet-15  & RecNet-20   \\ \midrule
%		\#Param    & 131,072  &  138,208   &   149,808   &   161,408   & \textbf{173,008}     \\
%		\#Flops & 131,072  & 46,268,416 & 103,940,096 & 161,611,776 & \textbf{219,283,456} \\
%		\#Memory (MB)    & 4,547  &  9,093   &  13,317   &  17,518   & \textbf{22,479}   \\ \bottomrule
%	\end{tabular}
%\end{table}

%图~\ref{fig:visual_MR_conv_kernels} 展示了训练完成的 RecNet-10 网络中部分卷积核可视化的结果。从中可以发现，浅层的卷积核学习到的特征多为基础部分，比如，边缘、线条等。中间层的卷积核能学习到组合的边、角、线等特征。后层的卷积核中，出现了更复杂的特征模式。这说明了经过大量的 MR 图像训练后，本章提出的重建网络 RecNet 能够很好的保留 MR 图像特有的特征规律，并且能恢复出MR图像的组织细节。
%
%\begin{figure}[th]
%	\centering
%	\includegraphics[scale=1.5]{visual_MR_conv_kernels.pdf}
%	\caption{本章提出的重建网络 RecNet-10 学习的部分卷积核可视化}
%	\label{fig:visual_MR_conv_kernels}
%\end{figure}

\subsection{Effect of Stable Constraints} \label{sec:effect_stable_constraint}

\begin{figure} [th]
	\centering
	\subfloat[30.05\% sampling rate with Eq.~\eqref{eq:mean_sample_rate_constraint}]{\includegraphics[width=0.45\linewidth]{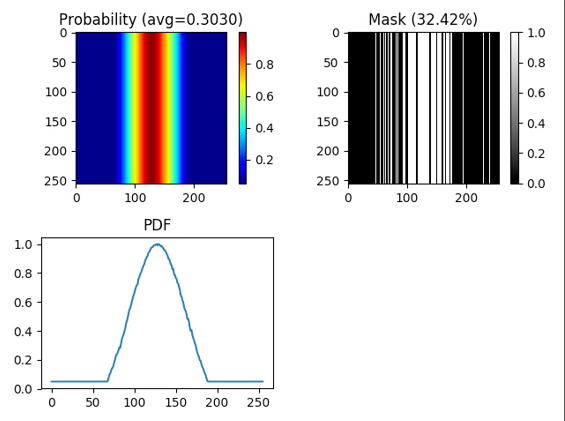}    	\label{subfig:probability_mask_10}} \hfil
%	\subfloat[10.05\% sampling rate with Eq.~\eqref{eq:mean_sample_rate_constraint}]{\includegraphics[width=0.42\linewidth]{probability_mask_10.pdf}    	\label{subfig:probability_mask_10}} \hfil
%	\subfloat[30.01\% sampling rate with Eq.~\eqref{eq:mean_sample_rate_constraint}]{\includegraphics[width=0.42\linewidth]{probability_mask_30.pdf}    	\label{subfig:probability_mask_30}} \\
%	\subfloat[10.05\% sampling rate with Eqs.~\eqref{eq:mean_sample_rate_constraint}--\eqref{eq:distance_constraint}]{\includegraphics[width=0.42\linewidth]{probability_mask_10_stable.pdf}    	
	\subfloat[50.05\% sampling rate with Eqs.~\eqref{eq:mean_sample_rate_constraint}--\eqref{eq:distance_constraint}]{\includegraphics[width=0.45\linewidth]{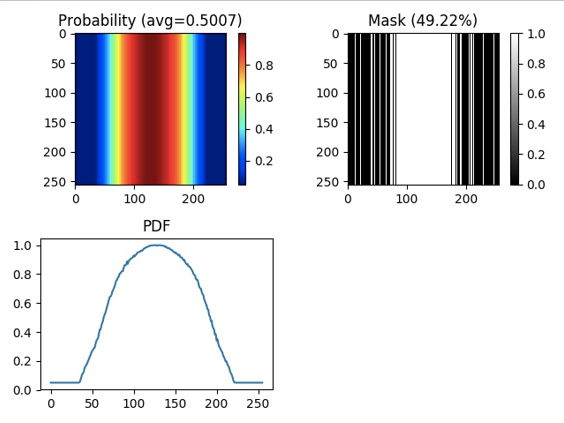}    
		\label{subfig:probability_mask_10_stable}} 
%	\subfloat[30.01\% sampling rate with Eqs.~\eqref{eq:mean_sample_rate_constraint}--\eqref{eq:distance_constraint}]{\includegraphics[width=0.42\linewidth]{probability_mask_30_stable.pdf} 
%		\label{subfig:probability_mask_30_stable}}
	\caption{Sampling matrices of our 1D probabilistic undersampling layer by sampling rate 10.05\% with Eq.~\eqref{eq:mean_sample_rate_constraint} and Eqs.~\eqref{eq:mean_sample_rate_constraint}--\eqref{eq:distance_constraint} respectively
	} \label{fig:probability_mask_stable_constraint}
\end{figure}

Fig.~\ref{fig:probability_mask_stable_constraint} shows the sampling matrices of our 1D probabilistic undersampling layer by sampling rate 10.05\% with Eq.~\eqref{eq:mean_sample_rate_constraint} and Eqs.~\eqref{eq:mean_sample_rate_constraint}--\eqref{eq:distance_constraint}, respectively. As shown in Fig.~\ref{subfig:probability_mask_10}, the sampling points in $ k $-space are distributed randomly since no regional sampling distance constraints~\eqref{eq:probability_constraint} and \eqref{eq:distance_constraint} are used. Especially in the high-frequency (HF) area, those sampling points generated by the Bernoulli distribution are either too sparse or too dense, though their probability values are the same. This apparently prevents us from analyzing its probability distribution. As shown in Fig.~\ref{subfig:probability_mask_10_stable}, those sampling points in $ k $-space are uniformly distributed by adopting \eqref{eq:probability_constraint} and \eqref{eq:distance_constraint}, especially in the HF area. This intuitively indicates that the probability values are equal in that area. In addition, after integrating the regional sampling distance constraint, the sampling matrices are quite stable and repeatable even through multiple trials. 

%图~\ref{fig:probability_mask_stable_constraint} 展示了采样比例为 10.05\% 和 30.01\% 时，本章分别加入总采样比例约束和区域采样间隔约束以克服随机性的干扰，训练获得的概率化降采样矩阵 (轨迹) 的可视化结果。由图~\ref{subfig:probability_mask_10} 和  \ref{subfig:probability_mask_30} 发现，未加入区域采样间隔约束的概率化降采样轨迹中，随机性使得$ k $空间采样点的位置间距变化剧烈。尤其是高频部分的区域，虽然它们的采样概率都一样，但是通过伯努利分布生成的采样点在有些区域很稀疏，而另一些区域很密集，这阻碍了本章直观上分析其概率分布的规律。由图~\ref{subfig:probability_mask_10_stable} 和  \ref{subfig:probability_mask_30_stable} 发现，加入区域采样间隔约束后，在训练得到的概率化降采样轨迹中，$ k $空间采样点的位置间距很均匀。尤其是高频区域，本章可以直观地解释这块区域的采样概率是相等的。另外，加入区域采样间隔约束后，概率化降采样轨迹的稳定性非常高，即使经过重复多次实验，得到的降采样轨迹基本一致。

\begin{figure}[t]
	\centering
%	\subfloat[]{\includegraphics[width=0.75\linewidth]{IFT_PSNR_vs_rate.pdf} 
%		\label{subfig:IFT_PSNR_vs_rate}} \\
%	\subfloat[]{\includegraphics[width=0.75\linewidth]{RecNet_PSNR_vs_rate.pdf}		
%		\label{subfig:RecNet_PSNR_vs_rate}} 
	\includegraphics[width=0.75\linewidth]{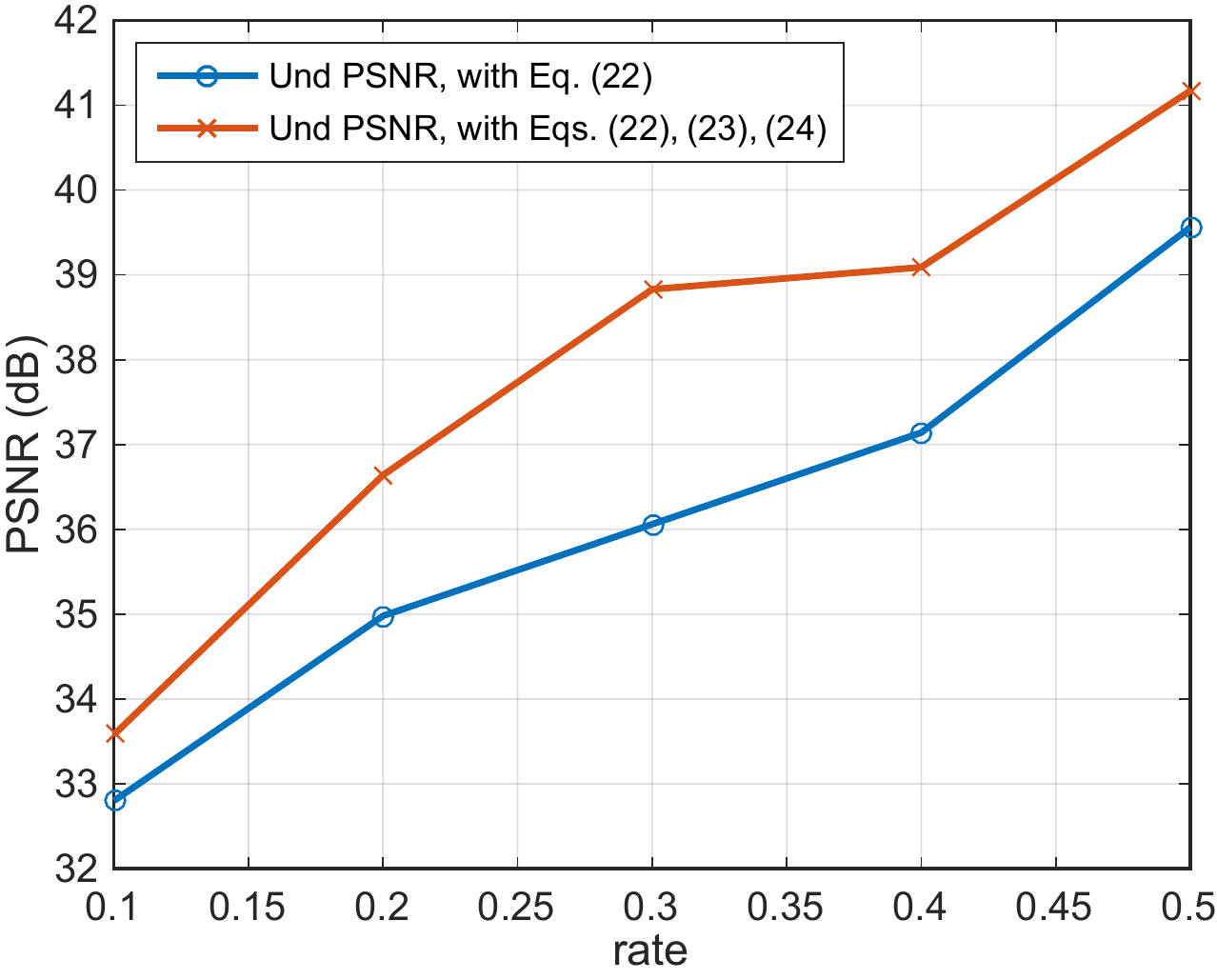}  \\
	\includegraphics[width=0.75\linewidth]{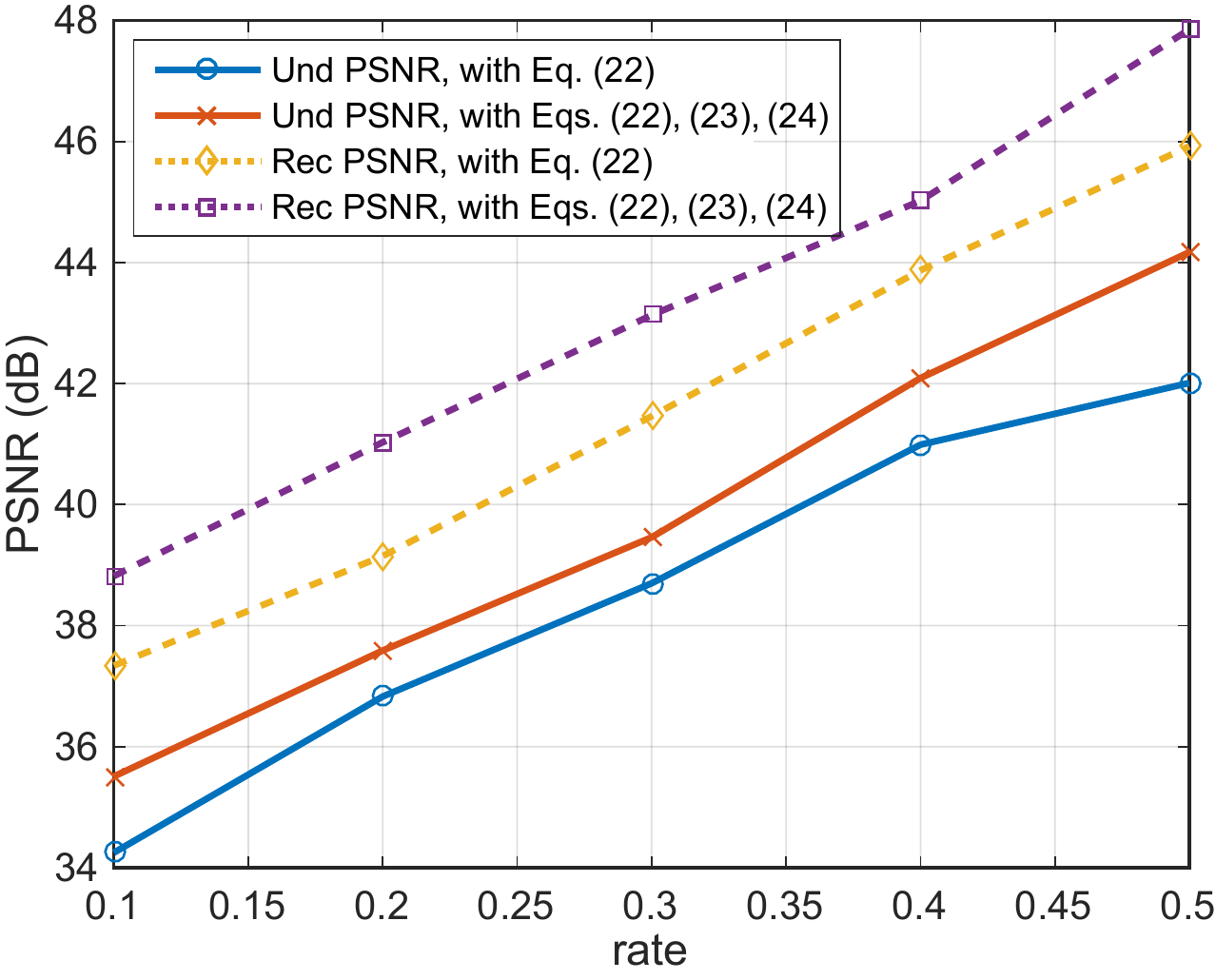}		
	\caption{PSNR of our cross-domain model  using the total sampling rate constraint only (Eq.~\eqref{eq:mean_sample_rate_constraint}) or together with the regional sampling distance constraint (Eqs.~\eqref{eq:mean_sample_rate_constraint}--\eqref{eq:distance_constraint}), respectively. The restuls were obtained from the Test3T\&7T set with sampling rate 10\%--50\%: (a) undersampling PSNR without RecNet; (b) undersampling / reconstruction PSNR with RecNet-10
%		降采样比例 10\%--50\% 条件下，本章在训练概率化降采样矩阵 (轨迹) 过程中分别加入总采样比例约束和区域采样间隔约束，并基于 Test3T\&7T 的 MR 图像集测试的结果
	}
	\label{fig:MR_RecNet_stable_constraint}
\end{figure}

%Table~\ref{tab:comparison_stable_reg_ift_PSNR} and 
Fig.~\ref{fig:MR_RecNet_stable_constraint}a shows the PSNR of our 1D probabilistic undersampling layer (no RecNet) using the total sampling rate constraint only or together with the regional sampling distance constraint, respectively. The results were based on the Test3T\&7T set with sampling rate 10\%--50\%. It can be seen that the regional sampling distance constraint not only helps generate stable sampling matrices, but also improves the undersampling PSNR of MR images by nearly 0.8--2.8~dB.
%表~\ref{tab:comparison_stable_reg_ift_PSNR} 和图~\ref{fig:MR_RecNet_stable_constraint}a 列出了降采样比例 10\%--50\% 条件下，无重建网络，本章在训练概率化降采样矩阵 (轨迹) 的过程中分别加入总采样比例约束和区域采样间隔约束，并基于 Test3T\&7T 的 MR 图像集测试的结果。可以发现，在总采样比例约束的前提下，区域采样间隔约束有助于生成稳定的降采样轨迹，并提升了降采样重建的效果。在多个采样比例条件下，它普遍能提升MR重建图像大约 0.8--2.8~dB 的 PSNR 指标。
%Table~\ref{tab:comparison_stable_reg_rec_PSNR} and 
Fig.~\ref{fig:MR_RecNet_stable_constraint}b shows the PSNR of our 1D probabilistic undersampling layer (with RecNet-10) using the total sampling rate constraint only or together with the regional sampling distance constraint, respectively. With the advantage of RecNet-10, our method can further improves the PSNR of final MR images by nearly 0.7--2.1~dB. Fig.~\ref{fig:MR_RecNet_stable_constraint} shows that the stable constraints and RecNet-10 both effectively improve the PNSR of our cross-domain model. 

%the improvement mainly results from the stable sampling matrices. Although RecNet-10 continually increases the values of PSNR, its effect has been quite limited. 

%表~\ref{tab:comparison_stable_reg_rec_PSNR} 和图~\ref{fig:MR_RecNet_stable_constraint}b 列出了降采样比例 10\%--50\% 条件下，结合重建网络 RecNet-10，本章在训练概率化降采样矩阵 (轨迹) 的过程中分别加入总采样比例约束和区域采样间隔约束，并基于 Test3T\&7T 的 MR 图像集测试的结果。可以发现，在总采样比例约束的前提下，区域采样间隔约束有助于生成稳定的降采样轨迹，并结合重建网络的参数，最大程度地提升MR图像的信噪比。在多个采样比例条件下，区域采样间隔约束能提升MR重建图像约 0.7--2.1~dB 的PSNR指标。另外，本章发现MR图像质量的提升主要依靠最佳的降采样轨迹，而训练后的RecNet-10虽然也能进一步提高 PSNR 值，但提升的效果已经非常有限。

The total sampling rate constraint ensures that the difference between the actual sampling rate and the target value is smaller than 0.1\%. This is convenient for us to analyze the relationship between the sampling matrix and the sampling rate. Thus, in the rest of experiments, we adopted the total sampling rate constraint and regional sampling distance constraint together (known as stable constraints), to obtain stable sampling matrices and substantially improve the PSNR of final MR images.

\subsection{Comparison of Different Undersampling Patterns} 
\label{sec:comparison_different_sampling_patterns}

\begin{table*}[t] \footnotesize
	\centering
	\caption{Comparison of PSNR by various undersampling methods (with RecNet-10). Results were based on Test3T\&7T set with sampling rate 10\%--50\%
%		降采样比例 10\%--50\% 条件下，结合重建网络 RecNet-10，本章提出的概率化降采样轨迹对比 J-CUR、PFGMS、J-MoDL、径向、高斯和泊松降采样轨迹，基于 Test3T\&7T 的 MR 图像集测试的结果。此处仅报告 PSNR，因为 SSIM 差别不大
}
	\label{tab:comparison_different_mask_rec_PSNR}
	\addtolength{\tabcolsep}{1pt}
	\begin{tabular}{cccccccc}
		\toprule
		Sampling &                                    \multicolumn{7}{c}{Undersampling PSNR (dB) / Reconstruction PSNR (dB)}                                    \\
		\cmidrule{2-8}
		rate &     J-CUR \cite{Weiss2020Joint}    &     PFGMS \cite{Gozcu2018LearningMRI}    &    J-MoDL \cite{Aggarwal2019J_MoDL}     &      LOUP \cite{Bahadir2019Learning}      &      Gaussian \cite{Cook1986Gaussian}      &      Poisson \cite{Jones2006Poisson}      &        {Probabilistic}   \\ \midrule
		10.09\%         & 27.23 / 30.85 & 29.51 / 31.02 & 30.19 / --.-- & 31.48 / 34.21 & 33.08 / 36.34 & 33.21 / 36.95 & \textbf{35.51 / --.--} \\
		20.01\%         & 29.46 / 31.34 & 31.35 / 33.24 & 32.37 / --.-- & 33.76 / 36.85 & 35.39 / 39.30 & 35.26 / 39.13 & \textbf{37.59 / --.--} \\
		29.91\%         & 30.54 / 32.83 & 32.48 / 35.91 & 33.89 / --.-- & 35.30 / 38.76 & 37.27 / 41.06 & 37.53 / 41.44 & \textbf{39.46 / --.--} \\
		40.04\%         & 31.09 / 33.15 & 33.72 / 37.67 & 35.64 / --.-- & 37.49 / 39.97 & 39.31 / 41.83 & 39.22 / 42.00 & \textbf{42.09 / --.--} \\
		50.00\%         & 32.64 / 34.79 & 35.83 / 39.67 & 37.28 / --.-- & 38.65 / 41.40 & 40.20 / 43.15 & 40.42 / 43.66 & \textbf{44.17 / --.--} \\ \bottomrule
	\end{tabular}
%	\vspace{1ex}
\end{table*}

Table~\ref{tab:comparison_different_mask_rec_PSNR} compares the PSNR values from different undersampling methods (with RecNet-10) based on Test3T\&7T set with sampling rate 10\%--50\%. As the sampling rate increases, the PSNR values of all methods improve gradually. It can be seen that the undersampling patterns of J-CUR and PFGMS perform worse than others, where the J-CUR only samples in 1D phase encoding direction and ignores the other dimension, as shown in Fig.~\ref{subfig:J_CUR_mask}. Since PFGMS allocates most sampling lines in the LF area, as shown in Fig.~\ref{subfig:PFGMS_mask}, its PSNR value is slightly better than J-CUR. Their common disadvantage is that the majority of HF information is lost, leading to low-quality of MR images. J-MoDL trains both the deep network and 1D undersampling pattern, but it samples evenly in the LF and HF areas, as shown in Fig.~\ref{subfig:JMoDL_mask}, so that its PSNR value is within 2.3~dB better than the two methods mentioned above. This indicates that the improvement of the undersampling pattern by J-MoDL is still limited. However, the undersampling pattern of LOUP, which is symmetric in Fig.~\ref{subfig:compared_LOUP_mask}, properly allocates the sampling points in LF and HF areas. Thus, it obviously improves the PSNR of recovered images. In addition, by using 1D probability functions, Gaussian and Poisson undersampling patterns have similar PSNR values and both outperform aforementioned methods. As a result, our proposed probabilistic undersampling pattern, integrated with RecNet-10, can obtain 2--3~dB higher values of PSNR than Gaussian and Poisson undersampling methods, under various sampling rates. Unlike other sampling patterns, our probability matrix can accurately control the status of each sampling point, and our sampling matrix only requires random initialization. By jointly cross-domain training the undersampling pattern and RecNet, our method can effectively obtain the best PSNR of the recovered MR images.

\begin{figure}[ht]
	\centering	
	\includegraphics[width=0.98\linewidth]{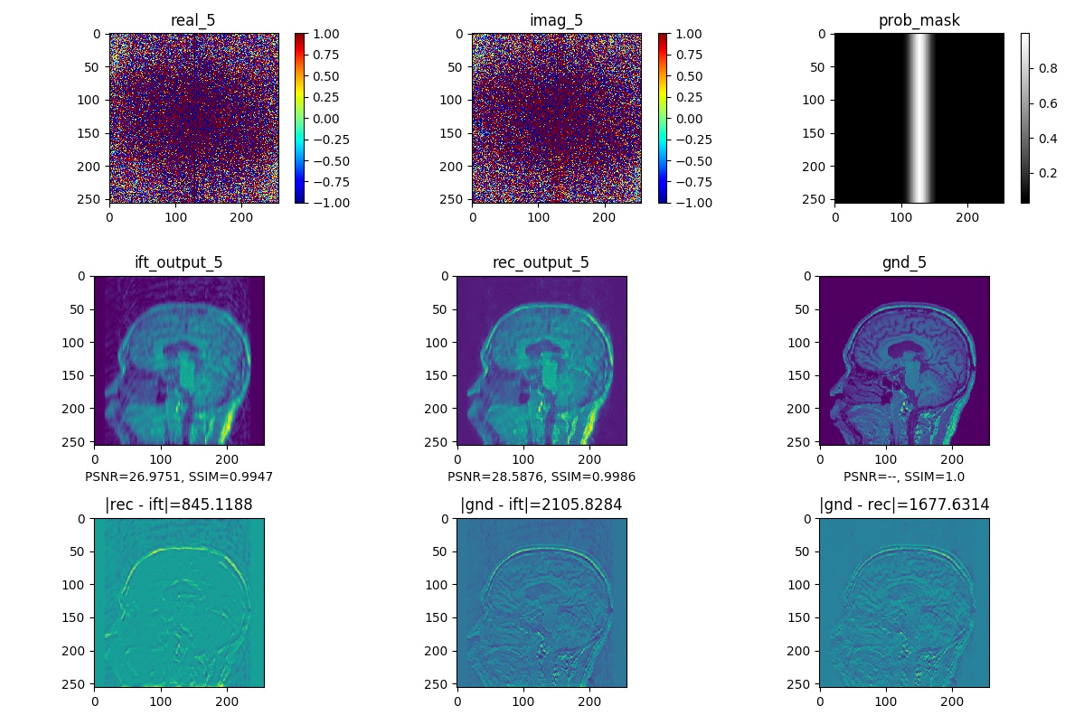}  \\ [1mm]	\includegraphics[width=0.98\linewidth]{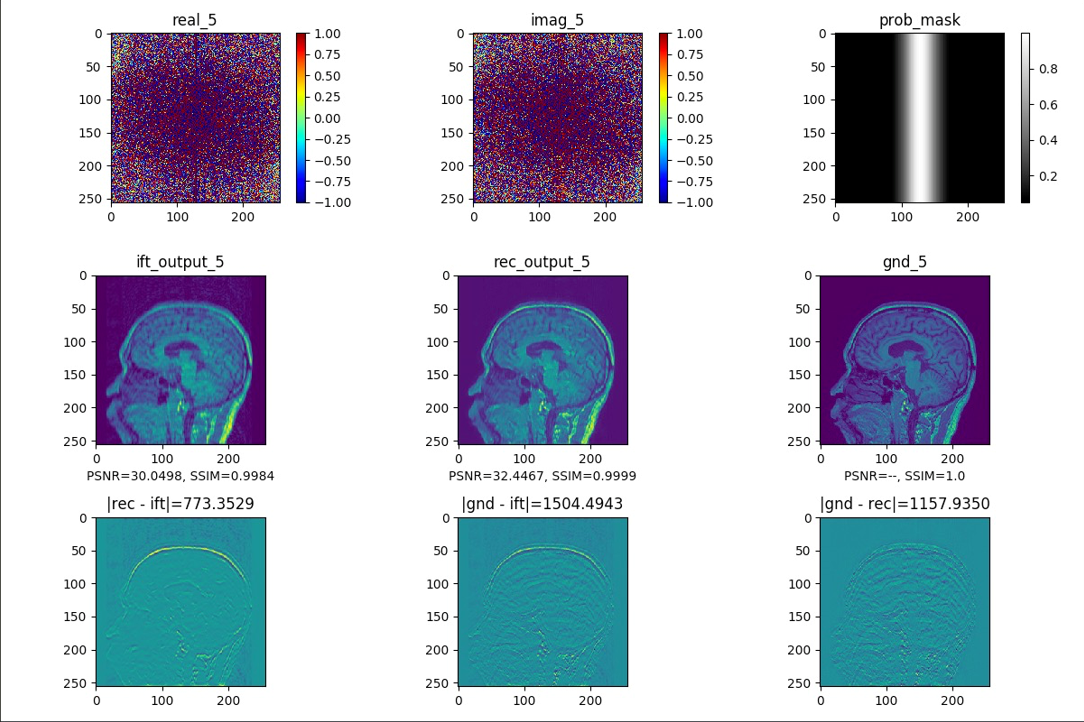}
	\caption{An example of comparison in recovered MR images resulting from different undersampling methods (with RecNet-10) based on the Test7T set with sampling rate 20\%: (a) Original; (b) J-CUR; (c) PFGMS; (d) J-MoDL; (e) LOUP; (f) Gaussian; (g) Poisson; (h) Probabilistic. Probabilistic undersampling method shows the best performance in reconstructing fine anatomic structures. The corresponding PSNR were labeled for each method.
%		降采样比例 20\% (5 倍加速)，结合重建网络 RecNet-10，本章提出的概率化降采样轨迹对比 J-CUR、PFGMS、J-MoDL、径向、高斯和泊松降采样轨迹，基于 Test7T 的 MR 图像集测试的结果: (a) 原始图像; (b) J-CUR 方法降采样; (c) PFGMS 方法降采样; (d) J-MoDL 方法降采样; (e) 径向降采样; (f) 高斯降采样; (g) 泊松降采样; (h) 概率化降采样
		%		{\color{red}{红色}}加粗数字表示最佳结果，{\color{blue}{蓝色}}\underline{下划线}数字表示次佳结果
	}
	\label{fig:RecNet_qualitative_comparison}
\end{figure}

Fig.~\ref{fig:RecNet_qualitative_comparison} shows an example of comparison in recovered MR images from different undersampling methods (with RecNet-10) based on the Test7T set with sampling rate 20\%. It can be seen that the recovered MR images by the J-CUR and PFGMS methods are quite blurry, and some details have been lost in the undersampling process. 
By increasing the sampling rate in the HF area, J-MoDL method obviously improves the PSNR of final results. However, its improvement is fairly limited, because J-MoDL deals with the LF and HF areas evenly, by allocating nearly identical sampling rates. 
In contrast, the LOUP method has most sampling points in the LF area and keeps a small number of sampling points randomly in the HF area, which apparently recovers small fine structures in MR images. 
In addition, by generating from 1D probability functions, the Gaussian and Poisson undersampling methods accurately distribute every sampling point in $ k $-space, which significantly reconstructs sophisticated structures and improves the quality of MR images. 
Our proposed 1D probabilistic undersampling pattern, trained with RecNet-10 on fully-sampled $ k $-space data, distributes  those sampling points in LF and HF areas by using the probability values. Thus, our method can effectively obtain higher PSNR values of MR images with more fine structures reconstructed than all other methods (red arrow in Fig.~\ref{fig:RecNet_qualitative_comparison}).
%Besides, we find that the improvement of PSNR mainly comes from our probabilistic undersampling pattern, which efficiently selects informative data in $ k $-space. After the undersampling process, RecNet-10 is then adopted to further improve the quality of MR images. 

%图~\ref{fig:RecNet_qualitative_comparison} 展示了在降采样比例 20\% (5 倍加速) 条件下，结合重建网络 RecNet-10，本章提出的概率化降采样轨迹对比 J-CUR、PFGMS、J-MoDL、径向、高斯和泊松降采样轨迹，基于 Test7T 的 MR 图像集测试的结果。
%可以发现，J-CUR 和 PFGMS 方法学习的降采样轨迹和重建网络的性能最差。它们输出的 MR 图像都相对模糊，许多边缘的细节已经在降采样的过程中丢失，即使利用重建网络也无法成功恢复。通过增加高频区域的采样比例，J-MoDL 方法学习的降采样轨迹以及重建网络提升了MR图像的信噪比，但是其对低频、高频区域分配了相同的采样比例，导致 PSNR 值的提升范围有限。相比而言，径向降采样轨迹明显协助了重建网络提取足量的高频信息，使得恢复的MR图像结构较完整，一些边缘部分已经变得清晰。
%借助1D概率分布，高斯降采样和泊松降采样轨迹结合了各自的重建网络，使得输出MR图像的 PSNR 指标出现显著的提升，因为其将低频、高频区域的采样点通过概率的形式进行较精确地分配，使得 MR图像中一些复杂的边缘结构已经被成功恢复。
%本章提出的概率化降采样轨迹，通过基于真实$ k $空间数据的跨域重建网络联合训练，得到了配对的一组采样矩阵和重建网络的参数。通过将低频、高频部分的采样点以独立同分布的概率值进行最合理地分配，本章获得了高信噪比的MR图像，使得图像中的组织结构大部分都被清晰地恢复。另外，本章要强调 PSNR 指标的提升主要依靠频域的最佳降采样轨迹，因为它实现了最有效地筛选$ k $空间的关键信息的功能。在获得欠采样的MR图像之后，重建网络RecNet-10的性能虽然很强大，但图像信噪比的提升空间已经非常有限。

\begin{figure}[ht]
	\centering	
	\includegraphics[width=0.98\linewidth]{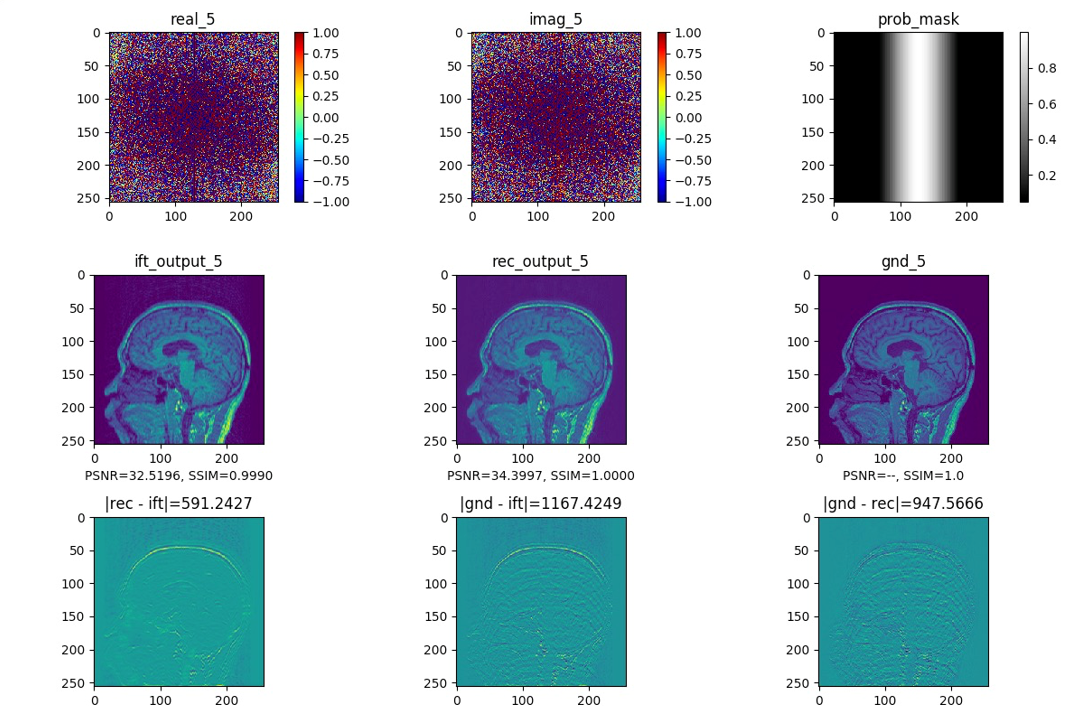}  \\ [1mm]	\includegraphics[width=0.98\linewidth]{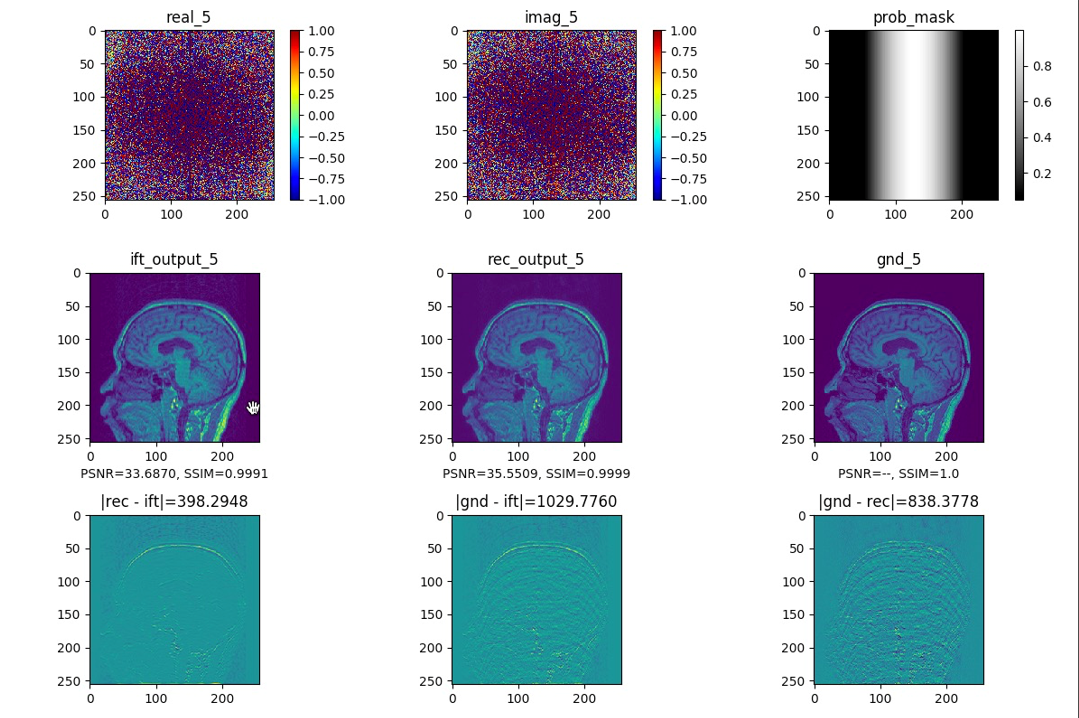}	
	\caption{An example of comparison in recovered MR images resulting from different undersampling methods (with RecNet-10) based on the Test7T set with sampling rate 20\%: (a) Original; (b) J-CUR; (c) PFGMS; (d) J-MoDL; (e) LOUP; (f) Gaussian; (g) Poisson; (h) Probabilistic. Probabilistic undersampling method shows the best performance in reconstructing fine anatomic structures. The corresponding PSNR were labeled for each method.
		%		降采样比例 20\% (5 倍加速)，结合重建网络 RecNet-10，本章提出的概率化降采样轨迹对比 J-CUR、PFGMS、J-MoDL、径向、高斯和泊松降采样轨迹，基于 Test7T 的 MR 图像集测试的结果: (a) 原始图像; (b) J-CUR 方法降采样; (c) PFGMS 方法降采样; (d) J-MoDL 方法降采样; (e) 径向降采样; (f) 高斯降采样; (g) 泊松降采样; (h) 概率化降采样
		%		{\color{red}{红色}}加粗数字表示最佳结果，{\color{blue}{蓝色}}\underline{下划线}数字表示次佳结果
	}
	\label{fig:RecNet_qualitative_comparison_2}
\end{figure}

\begin{figure}[ht]
	\centering	
	\includegraphics[width=0.98\linewidth]{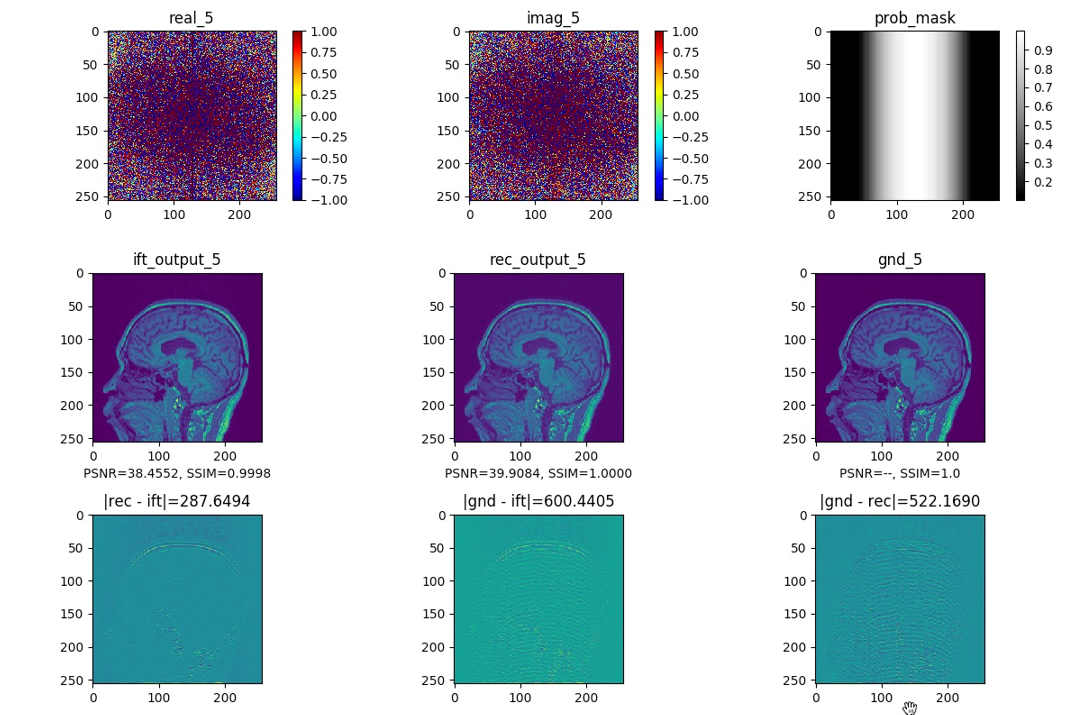} 
	\caption{An example of comparison in recovered MR images resulting from different undersampling methods (with RecNet-10) based on the Test7T set with sampling rate 20\%: (a) Original; (b) J-CUR; (c) PFGMS; (d) J-MoDL; (e) LOUP; (f) Gaussian; (g) Poisson; (h) Probabilistic. Probabilistic undersampling method shows the best performance in reconstructing fine anatomic structures. The corresponding PSNR were labeled for each method.
		%		降采样比例 20\% (5 倍加速)，结合重建网络 RecNet-10，本章提出的概率化降采样轨迹对比 J-CUR、PFGMS、J-MoDL、径向、高斯和泊松降采样轨迹，基于 Test7T 的 MR 图像集测试的结果: (a) 原始图像; (b) J-CUR 方法降采样; (c) PFGMS 方法降采样; (d) J-MoDL 方法降采样; (e) 径向降采样; (f) 高斯降采样; (g) 泊松降采样; (h) 概率化降采样
		%		{\color{red}{红色}}加粗数字表示最佳结果，{\color{blue}{蓝色}}\underline{下划线}数字表示次佳结果
	}
	\label{fig:RecNet_qualitative_comparison_3}
\end{figure}

\subsection{Analysis of 1D Probabilistic Undersampling Pattern}  \label{sec:analysis_optimal_undersampling_pattern}

To be continued ...

\section{Conclusion} \label{sec:conclusion}
%\section{Conclusion and Future work} \label{sec:conclusion}

In this paper, we have proposed a cross-domain network for MR image reconstruction, which contains a 1D probabilistic undersampling layer, a 1D inverse Fourier transform layer, and a reconstruction network. We can simultaneously obtain the optimal undersampling pattern (in \textit{k}-space) and the reconstruction model.
The 1D probabilistic undersampling layer, with differentiable probability and sampling matrices, can generate the optimal undersampling pattern and its probability distribution customized to specific $ k $-space data.
The 1D inverse Fourier transform layer connects the Fourier domain and the image domain during the forward pass and backpropagation.
By training 3D fully-sampled \textit{k}-space data and MR images with the conventional Euclidean loss, we have discovered the universal relationship between the probability function of the optimal undersampling pattern (sampling matrix) and its target sampling rate.
Retrospective experiments verify that the quality of recovered MR images by our 1D probabilistic undersampling pattern is apparently better than those by existing undersampling methods.

\setcounter{equation}{0}
\renewcommand{\theequation}{\thesection.\arabic{equation}}

\ifCLASSOPTIONcaptionsoff
  \newpage
\fi


\balance
\bibliographystyle{IEEEtran}
\bibliography{IEEEabrv, references}

\end{document}